# Mode fluctuations as fingerprint of chaotic and non-chaotic systems[*]

by

R. Aurich[†], A. Bäcker[‡] and F. Steiner[§]

Abteilung für Theoretische Physik, Universität Ulm
Albert-Einstein-Allee 11, D-89069 Ulm
Federal Republic of Germany

**Abstract:**
The mode-fluctuation distribution $P(W)$ is studied for chaotic as well as for non-chaotic quantum billiards. This statistic is discussed in the broader framework of the $E(k, L)$ functions being the probability of finding $k$ energy levels in a randomly chosen interval of length $L$, and the distribution of $n(L)$, where $n(L)$ is the number of levels in such an interval, and their cumulants $c_k(L)$. It is demonstrated that the cumulants provide a possible measure for the distinction between chaotic and non-chaotic systems. The vanishing of the normalized cumulants $C_k$, $k \geq 3$, implies a Gaussian behaviour of $P(W)$, which is realized in the case of chaotic systems, whereas non-chaotic systems display non-vanishing values for these cumulants leading to a non-Gaussian behaviour of $P(W)$. For some integrable systems there exist rigorous proofs of the non-Gaussian behaviour which are also discussed. Our numerical results and the rigorous results for integrable systems suggest that a clear fingerprint of chaotic systems is provided by a Gaussian distribution of the mode-fluctuation distribution $P(W)$.

[*]Supported by Deutsche Forschungsgemeinschaft under Contract No. DFG-Ste 241/6-1 and 7-1.
[†]E-mail address: aurich@physik.uni-ulm.de
[‡]E-mail address: baec@physik.uni-ulm.de
[§]E-mail address: steiner@physik.uni-ulm.de

# I  Introduction

One of the main research lines of quantum chaology is devoted to the search of fingerprints left on a quantum system by its classically chaotic counterpart. In classically bounded conservative systems, which are chaotic, one of the most obvious properties of chaos manifests itself in the long-time behaviour, like the exponential decay of certain correlation functions for $t \to \infty$. Their quantum mechanical counterparts do not display such a behaviour since the time-evolution operator $U(t) = e^{-\frac{i}{\hbar}Ht}$, $H$ being the Hamiltonian, leads to an almost periodic correlation function $\langle \phi | U(t) \phi' \rangle = \sum_{n=1}^{\infty} a_n e^{-\frac{i}{\hbar}E_n t}$, where $\phi, \phi'$ denote two states and $E_n$ the quantal levels of $H$. Thus, with respect to the long-time behaviour, there does not exist a fingerprint reflecting the classical properties. Therefore, in quantum chaology the limit $t = \infty$ is studied instead of the long-time asymptotics $t \to \infty$, which leads to the study of eigenstates, i.e., wave functions and quantal energy spectra, especially of their statistical properties.

This paper is devoted to a comparison between the statistics of the quantal levels of integrable and chaotic systems. For integrable systems, the short-range correlations in quantal energy spectra are, in general, numerically found to be Poissonian, but the long-range correlations show saturation effects [1, 2, 3]. In contrast, in chaotic systems the short-range correlations are numerically found to be in accordance with random-matrix theory (RMT) [4, 5, 6, 7], whereas long-range correlations experience modifications leading again to saturation effects. This general behaviour of chaotic systems is violated by the class of arithmetical systems [8, 9, 10, 11, 12, 13], where, e.g., the level-spacing distribution and the two-point statistics nearly behave as for classically integrable systems.

In most numerical studies concerning the statistical properties of quantal levels of chaotic systems, the nearest-neighbour level-spacing distribution $P(0, L)$, the number variance $\Sigma^2(L)$ or the spectral rigidity $\Delta_3(L)$ is studied. However, in this paper we will focus on the $E(k, L)$ functions, which give the probability to find $k$ levels in a randomly chosen interval of length $L$. They constitute a fundamental measure, since other statistics like the three just mentioned above are completely determined by them. Furthermore, their numerical calculation is simple and well defined, since no binning is necessary. In addition, as a function of $L$ one obtains smooth curves and significant results even in cases where only a small number of eigenvalues is known. The $E(k, L)$ functions are related to the moments and cumulants of the distribution of $n(L)$, which in turn are important for the behaviour of the mode-fluctuation distribution $P(W)$ [14, 15]. It now seems that the latter provides a clear fingerprint of quantum chaos.

The paper is organized as follows. In section II we provide some general remarks on the eigenvalue statistics of billiard systems, since we restrict our numerical analysis to those systems. Section III gives a summary of important properties of the $E(k, L)$ functions as well as of their behaviour in random matrix theory, and some phenomenological properties. Section IV discusses $n(L)$ and its corresponding distribution. The distribution of the normalized mode-fluctuations $W(x)$ and its connection to the distribution of $\eta_L(x)$ is discussed in section V. Exact results for mode-fluctuation distribution $P(W)$ in the case of integrable systems are summarized in section VI. Finally, section VII gives the numerical comparison between non-chaotic and chaotic systems.

## II General remarks on eigenvalue statistics in the case of quantum billiards

The systems to be studied are either given by the free motion of a point particle inside a compact Euclidean domain $\Omega$ with elastic reflections at the boundary $\partial\Omega$, or by the geodesic motion on a compact Riemann surface with constant negative curvature, or by billiards on compact domains with constant negative curvature. In each case the quantum mechanical system is determined by the stationary Schrödinger equation ($\hbar = 2m = 1$)

$$-\Delta \Psi_n(\mathbf{q}) = E_n \Psi_n(\mathbf{q}) \ , \tag{1}$$

with the appropriate Dirichlet, Neumann or periodic boundary conditions, respectively. $\Delta$ denotes the Laplace-Beltrami operator, which on the Euclidean plane is the usual Laplacian. Billiard systems belong to the class of scaling systems, which means that one has the scaling property $S_\gamma(\lambda E) = \lambda^\alpha S_\gamma(E)$ of the action $S_\gamma(E) = \int_\gamma \mathbf{p} \, d\mathbf{q}$ of periodic orbits $\gamma$. If one defines $\tilde{x} = E^\alpha$, one has $S_\gamma(E) = (E/E_0)^\alpha S_\gamma(E_0) =: \tilde{x} R_\gamma$ where $R_\gamma$ is independent of $E$. Also geodesic flows on Riemannian manifolds are scaling systems, where $\tilde{x} = p = \sqrt{E - 1/4}$ is the modulus of the momentum. Hamiltonian systems with potentials scaling according to $V(\lambda \mathbf{q}) = \lambda^\kappa V(\mathbf{q})$, $\lambda > 0$, provide another class of scaling systems, where $\alpha = \frac{1}{2} + \frac{1}{\kappa}$. For billiards the semiclassical limit corresponds to the high-energy limit $E_n \to \infty$.

Due to the compactness of $\Omega$, the quantal energy spectrum $\{E_n\}$ is purely discrete; thus one can define the spectral staircase function $N(E)$ (mode number, integrated level density)

$$N(E) := \#\{n | E_n \leq E\} \ , \tag{2}$$

which counts the number of energy levels $E_n$ below a given energy $E$. $N(E)$ can be separated into a mean smooth part $\overline{N}(E)$ and a fluctuating part, the *mode fluctuations* $N_{\text{fluc}}(E)$

$$N(E) = \overline{N}(E) + N_{\text{fluc}}(E) \ . \tag{3}$$

For two-dimensional billiards, $\overline{N}(E)$ is given by the generalized Weyl's law [16] and an additional term $N_{\text{long}}(E)$

$$\overline{N}(E) = \frac{\mathcal{A}}{4\pi} E - \frac{\mathcal{L}}{4\pi} \sqrt{E} + \mathcal{C} + N_{\text{long}}(E) \ , \tag{4}$$

where $\mathcal{A}$ denotes the area of the billiard, and $\mathcal{L} := \mathcal{L}^- - \mathcal{L}^+$, where $\mathcal{L}^-$ and $\mathcal{L}^+$ are the lengths of the boundary $\partial\Omega$ with Dirichlet and Neumann boundary conditions, respectively. $\mathcal{C}$ takes curvature and corner corrections into account. The term $N_{\text{long}}(E)$ is zero or negligible for most systems, but for some systems $N(E)$ possesses a slowly varying long-range oscillation which we denote by $N_{\text{long}}(E)$. This was first observed [17] in the case of the stadium billiard [18, 19] (see sec. VII.2), where by subtracting the contribution of the so-called bouncing-ball orbits [20] the expected behaviour of the spectral statistics could be restored. Further contributions to $N_{\text{long}}(E)$ in the case of the stadium billiard are higher order corrections [20, 21] caused by edge-orbits, which run into the point where circle and straight line join. The same applies to the truncated hyperbola billiard (see section VII.2), where a prominent contribution to $N(E)$ is given by families of closed non-periodic orbits running into a boundary point where the curvature is discontinuous [22]; but here their contribution vanishes for $E \to \infty$, whereas the contribution of the bouncing-ball orbits in the stadium billiard increases in an unbounded manner. The main difference between $N_{\text{long}}(E)$ and $N_{\text{fluc}}(E)$ is that the oscillations in $N_{\text{long}}(E)$ are regular whereas the fluctuations in $N_{\text{fluc}}(E)$ are of an irregular nature. These irregular fluctuations are exactly those in which one is interested in the theory of quantum chaos.

To compare the quantal spectra of different systems, one has to get rid off the system-dependent constants in $\overline{N}(E)$, which is achieved by *unfolding* the spectra [23] by

$$x_n := \overline{N}(E_n) \quad , \quad n = 1, 2, 3, \ldots \quad . \tag{5}$$

The unfolded spectrum $\{x_n\}$ has by construction a unit mean level spacing. In the sequel we will assume that the spectra are already unfolded, and in order to keep notation simple, $N(x)$ will denote the spectral staircase of the unfolded energy spectrum, and $N_{\text{fluc}}(x)$ the corresponding fluctuating part. Thus $N(x) = x + N_{\text{fluc}}(x)$.

To investigate the statistical distribution of the spectrum, several statistical measures have been introduced [23, 24]. The simplest quantity is $\delta_n$, which is obtained from the mode fluctuations, i.e., the fluctuating part of the spectral staircase evaluated at the unfolded energy eigenvalues $x_n$

$$\delta_n := N_0(x_n) - \overline{N}(x_n) = n - \frac{1}{2} - x_n \quad , \tag{6}$$

where $N_0(x) := \lim_{\epsilon \to 0} \frac{N(x+\epsilon)+N(x-\epsilon)}{2}$. The quantity $\delta_n$ can serve as a good measure for the completeness of a given energy spectrum.

Another statistic is the level-spacing distribution $P(k, L)$ which is the probability density of the distances $x_{n+k+1} - x_n$ between energy levels. The well-known nearest-neighbour level-spacing is obtained for $k = 0$, $P(L) = P(0, L)$.

The number variance allows one to investigate medium- and long-range correlations of the spectrum. It is defined as the local variance of the number $n(L, x) := N(x + L) - N(x)$ of unfolded energy-levels in the interval $[x, x + L]$

$$\Sigma^2(L, \hat{x}) := \left\langle [n(L, x) - L]^2 \right\rangle_{\hat{x}} \quad , \quad L > 0 \quad . \tag{7}$$

For the averaging there are a priori different possibilities. We choose the following rectangular averaging

$$\langle f(x) \rangle_{\hat{x}} = \frac{1}{(c-1)\hat{x}} \int_{\hat{x}}^{c\hat{x}} f(x) \, dx \quad , \tag{8}$$

where $c > 1$. The expectation value of $n(L, x)$ is given by

$$\langle n(L, x) \rangle_{\hat{x}} = \frac{1}{(c-1)\hat{x}} \int_{\hat{x}}^{c\hat{x}} \{ N(x + L) - N(x) \} \, dx$$

$$= L + \frac{1}{(c-1)\hat{x}} \int_{\hat{x}}^{c\hat{x}} \{ N_{\text{fluc}}(x + L) - N_{\text{fluc}}(x) \} \, dx \quad ,$$

which tends to $L$ in the semiclassical limit of $\hat{x} \to \infty$, since $N_{\text{fluc}}(x)$ has zero mean.

A related statistic, which also measures two-point correlations, is the spectral rigidity [25]

$$\Delta_3(L, \hat{x}) := \left\langle \min_{(a,b)} \frac{1}{L} \int_{-\frac{L}{2}}^{\frac{L}{2}} dy \, [N(x + y) - a - by]^2 \right\rangle_{\hat{x}} \quad , \tag{9}$$

which is the average mean square deviation of the spectral staircase function from the best fitting straight line over an energy range of $L$ mean level spacings.

It has been conjectured that the energy level statistics of integrable systems can be described by a Poissonian random process [26], whereas the statistics of classically strongly chaotic systems should be reproduced by the corresponding RMT distributions of the GOE or GUE [4, 5, 6]. It turned out that this is only true for short-range correlations, e.g., measured by the nearest-neighbour level-spacing distribution $P(0, L)$ [1, 27]. A convenient measure for the long-range correlations is provided by the number variance $\Sigma^2(L, \hat{x})$, which, however, does not discriminate the different short-range behaviours since one gets the behaviour $\Sigma^2(L, \hat{x}) = L + O(L^2)$ solely from the fact that the spectra are unfolded. The RMT results for large $L$ are $\Sigma^2_{\text{GOE}}(L) \simeq \frac{2}{\pi^2} \ln(2\pi L)$ and $\Sigma^2_{\text{GUE}}(L) \simeq \frac{1}{\pi^2} \ln(2\pi L)$ which is in strong contrast to the observed saturation of the number variance (see, e.g., [7, 11, 28, 29, 30]). The first chaotic system for which the saturation behaviour was studied was the Hadamard-Gutzwiller ensemble [31]. Berry has given semiclassical arguments in favour of the saturation behaviour of the number variance [30, 32] and of the spectral rigidity [2]. It turned out that with $L_{\max} := a\sqrt{\hat{x}}$, where $a > 0$ is a system-dependent constant, one has the universality regime for $L \ll L_{\max}$, in which the spectral statistics are expected to be described by random matrix theory. For $L \gg L_{\max}$ the number variance oscillates around its saturation plateau $\Sigma^2_{\infty}(\hat{x})$. This saturation plateau increases with increasing energy $\hat{x}$ in the limit of large $\hat{x}$.

## III $E(k, L)$ functions

The $E(k, L)$ functions, $k = 0, 1, 2, \ldots$, are defined as the probabilities to find $k$ levels in a randomly chosen interval of length $L$ [25, 33]. In RMT the $E(k, L)$ functions do not depend on the considered interval. For real systems, however, this stationarity is not fulfilled. Therefore, one considers $E(k, L, \hat{x})$, the probability of $n(L, x)$ being equal to $k$, where $x$ is chosen uniformly distributed in the interval $x \in [\hat{x}, c\hat{x}]$, i.e.,

$$E(k, L, \hat{x}) = \frac{1}{(c-1)\hat{x}} \int_{\hat{x}}^{c\hat{x}} \delta_{k, n(L, x)} \, dx \quad , \tag{10}$$

where $\delta_{n,m}$ is the Kronecker symbol. Since $E(k, L, \hat{x})$ and $\Sigma^2(L, \hat{x})$ are only directly related if one chooses the same averaging in both statistics, we consistently use the rectangular averaging (8). In order to keep the notation simple, we will write $E(k, L)$ and $\Sigma^2(L)$.

From the definition follows that the $E(k, L)$ functions have their maximum approximately at $L = k$ since only unfolded spectra are considered. Furthermore, they satisfy the normalization

$$\sum_{k=0}^{\infty} E(k, L) = 1 \quad . \tag{11}$$

The number of levels $n(L)$ in an interval of length $L$ has the expectation value

$$\langle n(L) \rangle_{\hat{x}} = \sum_{k=0}^{\infty} k \, E(k, L) \simeq L \quad . \tag{12}$$

The last asymptotic equality holds strictly in the case of RMT.

The $E(k, L)$ functions determine the other statistics as we would like to illustrate in some examples. Denoting by $F(k, L)$ the probability to count $k$ levels in an interval of length $L$, which starts at an arbitrary quantal level $x_i$, one has the relation

$$F(k, L) = -\frac{\partial}{\partial L} \sum_{i=0}^{k} E(i, L) \quad . \tag{13}$$

The level-spacing distribution $P(k, L)$, which is the probability to obtain exactly $k$ levels in an interval of length $L$ beginning at an arbitrary quantal level $x_i$ and ending at the quantal level $x_{i+k+1}$, is given by

$$P(k, L) = -\frac{\partial}{\partial L} \sum_{i=0}^{k} F(i, L) = \frac{\partial^2}{\partial L^2} \sum_{i=0}^{k} (k - i + 1) E(i, L) \quad . \tag{14}$$

The nearest-neighbour level-spacing is the special case $k = 0$. As a last example we would like to recall that the number variance $\Sigma^2(L)$ is given by

$$\Sigma^2(L) = \sum_{k=0}^{\infty} (k - L)^2 E(k, L) \quad . \tag{15}$$

In the case of a Poissonian random process one has an analytic expression for the $E(k, L)$ functions

$$E_{\text{Poisson}}(k, L) = \frac{1}{k!} L^k e^{-L} \quad . \tag{16}$$

In figure 1 the $E_{\text{Poisson}}(k, L)$ functions are shown in comparison with the $E(k, L)$ functions obtained from the first 100 000 quantal levels of the circular billiard being an example of an integrable system (see section VII). For small values of $L$ there is quite good agreement of the $E(k, L)$ functions with $E_{\text{Poisson}}(k, L)$, but for larger values of $L$ clear deviations are visible which reflect the saturation of the $E(k, L)$ functions.

## III.1    RMT behaviour for the $E_\beta(k, L)$ functions

RMT starts with the joint probability measure $\mathcal{P}_\beta(x_1, \ldots, x_N) \, dx_1 \ldots dx_N$ [25, 34], which is the probability of finding a level in each of the intervals $[x_i, x_i + dx_i]$, $i = 1, \ldots, N$, regardless of the labelling, i.e., $\mathcal{P}_\beta(x_1, \ldots, x_N)$ is symmetric under the interchange of $x_i$ with $x_j$. For the two commonly used ensembles, the Gaussian orthogonal ensemble (GOE) and the Gaussian unitary ensemble (GUE), RMT yields [25]

$$\mathcal{P}_\beta(x_1, \ldots, x_N) = C_\beta \exp\left\{-\frac{\beta}{2} \sum_{i=1}^{N} x_i^2\right\} \prod_{i<j} |x_i - x_j|^\beta \tag{17}$$

with $\beta = 1$ for the GOE and $\beta = 2$ for the GUE, where $C_\beta$ is a normalization constant. Since RMT deals with matrices of infinite dimension the limit $N \to \infty$ has to be considered.

The $E(k, L)$ functions can be obtained from $\mathcal{P}_\beta(x_1, \ldots, x_N)$ in the following way. To get exactly $k$ levels in an interval $I = [X, X + L]$ one has to integrate $\mathcal{P}_\beta(x_1, \ldots, x_N)$ for the first $k$ levels over the interval $I$ and the remaining levels over the range excluding $I$

$$E_\beta(k, L) = \frac{N!}{(N - k)!} \int_I dx_1 \ldots dx_k \int_{\mathbf{R} \setminus I} dx_{k+1} \ldots dx_N \, \mathcal{P}_\beta(x_1, \ldots, x_N) \quad , \tag{18}$$

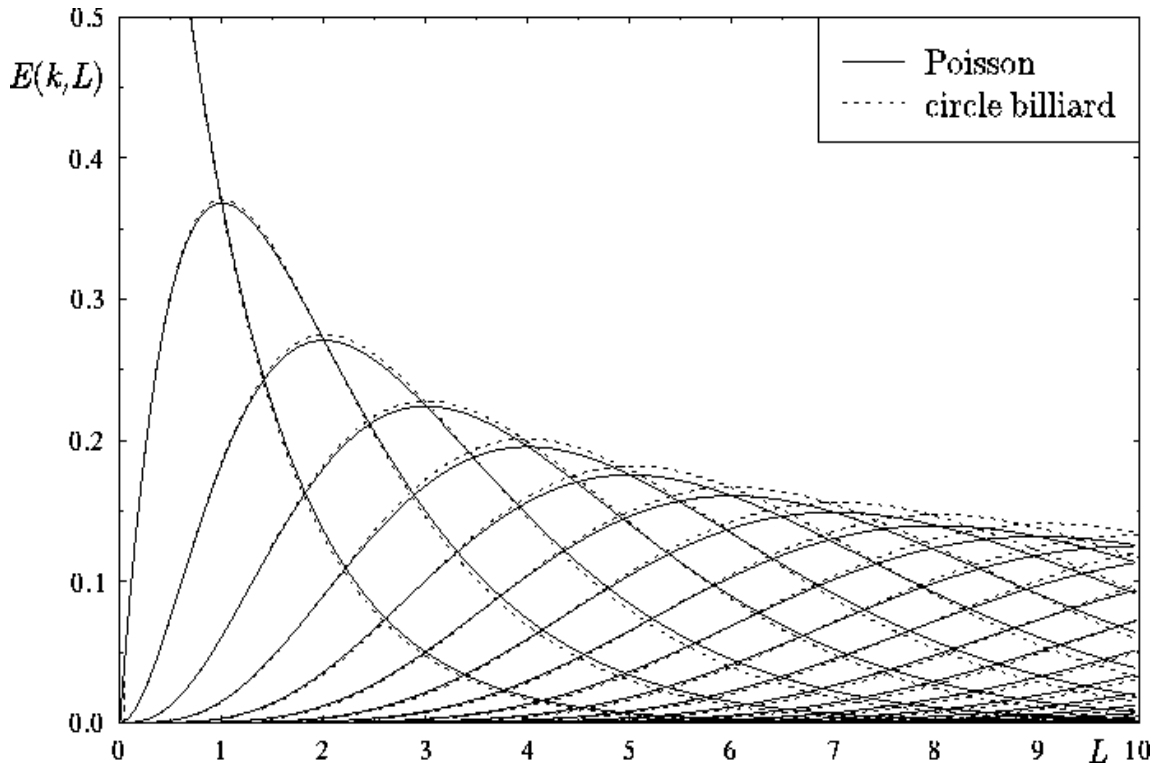

Figure 1: The $E(k,L)$ functions are shown for the Poisson distribution and for the circular billiard discussed in section VII where the quantal levels up to the 100 000th are used. For a given $k$, the corresponding $E(k,L)$ function is given by the curve having its maximum at $L \approx k$.

where the combinatoric factor is due to the symmetry with respect to the interchange of $x_i$ with $x_j$. This function depends on the choice of the interval $I$, i.e., on $X$. If, however, the spectrum is stationary, i.e., invariant under translations, as it is the case for GOE and GUE [35, 36], the result is independent of this choice after $N \to \infty$. For the systems which will be discussed later an energy dependence is observed, and thus the results depend on the choice of $I$, i.e., on $X$.

This integration process leads to the study of the continuous, real, symmetric integral kernels [25, 33, 37]

$$K(x,y) := \frac{\sin\left(\frac{\pi L}{2}(x-y)\right)}{\pi(x-y)} \quad \text{and} \quad K_{\pm}(x,y) := K(x,y) \pm K(x,-y) \quad , \quad x,y \in [-1,1] \quad , \tag{19}$$

since the $E_{\beta}(k,L)$ are related to the Fredholm determinants $\mathcal{D}(L,\gamma) = \det(1 - \gamma K)$ and $\mathcal{D}_{\pm}(L,\gamma) = \det(1 - \gamma K_{\pm})$. For the GUE case ($\beta = 2$) one obtains

$$E_2(k,L) = \frac{(-1)^k}{k!} \left.\frac{\partial^k \mathcal{D}(L,\gamma)}{\partial \gamma^k}\right|_{\gamma=1} \quad . \tag{20}$$

Thus the Fredholm determinant $\mathcal{D}(L,\gamma)$ can be expressed as

$$\mathcal{D}(L,\gamma) = \sum_{k=0}^{\infty} (1-\gamma)^k E_2(k,L) \quad . \tag{21}$$

The GOE case ($\beta = 1$) is obtained by defining analogously to (20) functions $E_\pm(k, L)$ from which the $E_1(k, L)$ are determined by the relations

$$
\begin{aligned}
E_+(k, L) &= E_1(2k, L) + E_1(2k - 1, L) \quad, \quad k > 0 \\
E_-(k, L) &= E_1(2k, L) + E_1(2k + 1, L) \quad, \quad k \geq 0
\end{aligned}
\tag{22}
$$

with $E_1(0, L) = E_+(0, L)$. The Fredholm determinants are computed by diagonalization of the integral kernel $K(x, y)$ leading to the problem of solving the eigenvalue problem

$$
\int_{-1}^{1} dy\, K(x, y)\, f_i(y, L) = \lambda_i(L)\, f_i(x, L) \quad,
\tag{23}
$$

where the eigenfunctions $f_i(x, L)$ are given by the prolate spheroidal functions. They possess a definite parity $f_i(x, L) = \pm f_i(-x, L)$ because of the parity symmetry $(x, y) \to (-x, -y)$ of the integral kernel. The Fredholm determinants are then given by the products

$$
\mathcal{D}(L, \gamma) = \prod_{i=0}^{\infty}(1 - \gamma \lambda_i(L))
\tag{24}
$$

and

$$
\mathcal{D}_+(L, \gamma) = \prod_{i=0}^{\infty}(1 - \gamma \lambda_{2i}(L)) \quad, \quad \mathcal{D}_-(L, \gamma) = \prod_{i=1}^{\infty}(1 - \gamma \lambda_{2i-1}(L)) \quad.
\tag{25}
$$

With the above equations it is possible to compute the $E_\beta(k, L)$ alone from the eigenvalues $\lambda_i(L)$ of the prolate spheroidal functions. It is worthwhile to note that the original version provided in [33] requires the computation of integrals over the prolate spheroidal functions $f_i(x, L)$, a procedure which is now obsolete. The differentiation has to be carried out analytically to obtain numerically stable results. For the GUE case one gets

$$
E_2(k, L) = \sum_{i_1 < \ldots < i_k} \lambda_{i_1}(L) \ldots \lambda_{i_k}(L) \prod_{\substack{j=0 \\ j \notin \{i_1, \ldots, i_k\}}}^{\infty} (1 - \lambda_j(L)) \quad.
\tag{26}
$$

For the GOE case one has similar expressions for $E_+(k, L)$ and $E_-(k, L)$ where only even and odd eigenvalues occur, respectively. The $E_1(k, L)$ are then given by (22). Although the product runs over infinitely many eigenvalues, a fast convergence is numerically observed since almost all eigenvalues are very close to zero or to one. The above representation of the $E(k, L)$ without the integrals over the spheroidal functions has also valuable applications as the determination of power series in $L$ [37, 38] or the asymptotic behaviour $L \to \infty$ of the $E_\beta(k, L)$ [39], which is also investigated by other methods in [40, 41].

Let us now turn to the integral equation (23) which will be solved separately for even and odd parity. The prolate spheroidal functions are expanded in Legendre polynomials with fixed parity up to order $2N$ or $2N + 1$, i.e., for even parity $f_i(x, L) = \sum_{k=0}^{N} a_k(L)\, P_{2k}(x)$, and for odd parity $f_i(x, L) = \sum_{k=0}^{N} a_k(L)\, P_{2k+1}(x)$. To discretise the eigenvalue problem, $N + 1$ points $x_i$ in the interval $[0, 1]$ are chosen which provide $N + 1$ equations by inserting these points into (23). Let $\vec{a} := (a_k)$, then the generalized eigenvalue problem $A\vec{a} = \lambda B\vec{a}$ is obtained whose matrix elements for even parity read

$$
A_{jk} = \int_{-1}^{1} dy\, K(x_j, y)\, P_{2k}(y) \quad \text{and} \quad B_{jk} = P_{2k}(x_j) \quad.
\tag{27}
$$

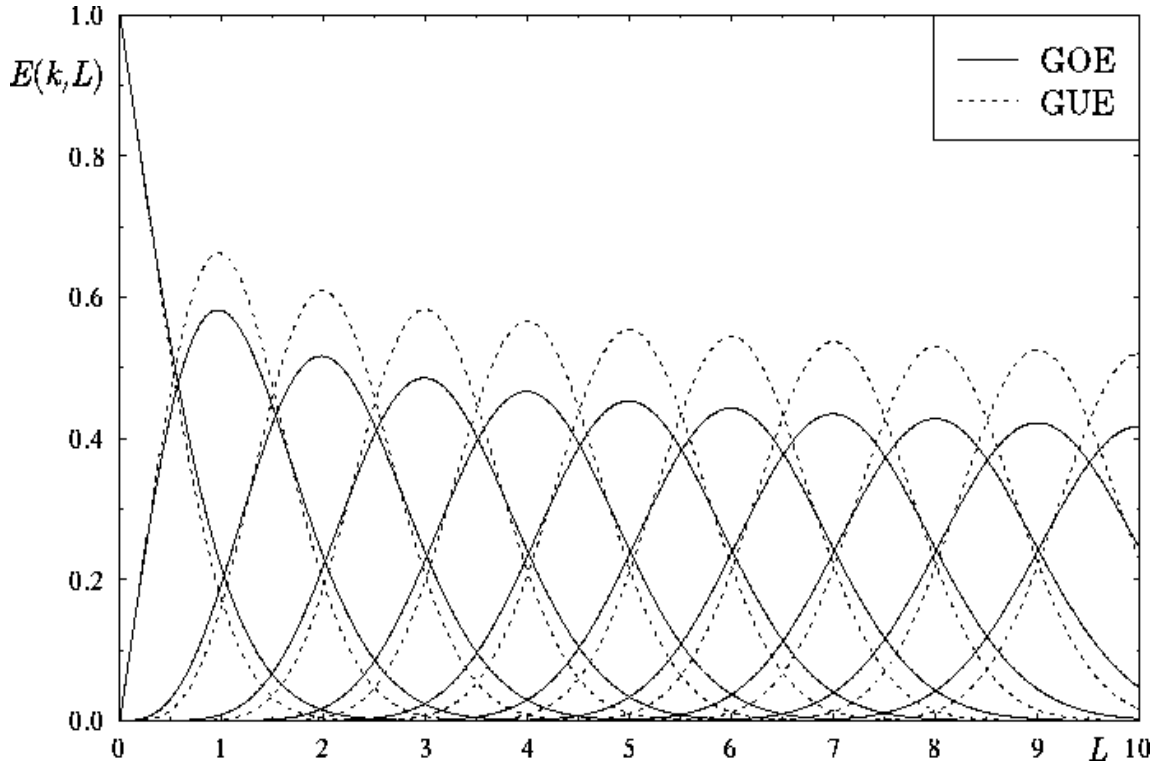

Figure 2: The $E(k, L)$ functions are shown for the GOE (full curves) and for the GUE case (dotted curves).

Analogous matrix elements yield the eigenvalue problem for odd parity. The numerical solution of the eigenvalue problem provides the eigenvalues $\lambda_i(L)$ as well as the prolate spheroidal functions, which are determined by $\vec{a}$. We choose $N = 20, \ldots, 25$ which allows sufficiently accurate results. In figure 2 the $E(k, L)$ functions are displayed for the GOE and the GUE case.

Often in the statistical analysis of energy spectra, main emphasis is laid on the investigation of short-range correlations by considering, e.g., the nearest-neighbour level-spacing distribution $P(0, L)$. The "corresponding" $E(k, L)$ function is $E(0, L)$ which is the probability of finding no eigenvalue in an interval of length $L$, and thus it is also called *gap-probability*. In this case good approximations for $E(0, L)$ can be obtained using (c.f. equation (14))

$$P(0, L) = \frac{\partial^2}{\partial L^2} E(0, L) \qquad (28)$$

from the approximations to $P(0, L)$: In the case of GOE the nearest-neighbour level-spacing distribution $P(0, L)$ is very well described by the Wigner distribution

$$P^{\text{GOE}}(0, L) \simeq \frac{\pi}{2} L \exp\left\{-\frac{\pi}{4} L^2\right\} \quad, \qquad (29)$$

whereas for GUE one has

$$P^{\text{GUE}}(0, L) \simeq \frac{32}{\pi^2} L^2 \exp\left\{-\frac{4}{\pi} L^2\right\} \quad. \qquad (30)$$

Using (29) and (30) in equation (28), one calculates $E(0, L)$ by integrating twice. The integration constants are determined by the fact that one has to fulfill $E(0, 0) = 1$ and $\lim_{L \to \infty} E(0, L) = 0$.

One obtains

$$E_1(0, L) \simeq \text{erfc}\left(\frac{\sqrt{\pi}}{2}L\right) \qquad (31)$$

$$E_2(0, L) \simeq \exp\left(-\frac{4}{\pi}L^2\right) - L\,\text{erfc}\left(\frac{2}{\sqrt{\pi}}L\right) \ . \qquad (32)$$

Remarkably, the approximations for $E_1(0, L)$ and $E_2(0, L)$ are rather accurate (absolute error less than 0.005), despite the fact that they are based on approximations, which then were integrated twice.

## III.2   Phenomenological description of the $E(k, L)$ functions

In the case of GOE and GUE [31] as well as for classically chaotic systems [31, 42] it is found that the $E(k, L)$ functions are excellently approximated for some $k \geq k_0$ by

$$E(k, L) \simeq \frac{1}{\sqrt{2\pi\alpha(L)}}\exp\left(-\frac{(L-k)^2}{2\alpha(L)}\right) \ . \qquad (33)$$

The relation of the width $\alpha(L)$ to the number variance $\Sigma^2(L)$ [31],

$$\alpha(L) = \Sigma^2(L) \ , \quad L \gg k_0 \ , \qquad (34)$$

is obtained by inserting the approximation (33) for $k \geq k_0$ in eq.(15) and by approximating the sum by the Euler-MacLaurin summation formula. A numerical comparison shows that the $E_\beta(k, L)$ functions belonging to GOE and GUE are well described by (33) already for $k_0 = k_{\text{RMT}} \simeq 6$. Since for chaotic systems it is observed that the $E(k, L)$ functions agree with the $E_\beta(k, L)$ functions for $k \ll k_s \sim L_{\max}$ with $k_s$ increasing with energy, one gets for chaotic systems $k_0 \simeq 6$, in general.

The approximation (33) describes a Gaussian distribution with width $\alpha(L)$ if the $E(k, L)$ functions are considered as a function of $k$ with $L$ fixed. However, this is a discrete statistic since $k$ is an integer. On the other hand, if $\alpha(L) = \Sigma^2(L)$ is only slowly varying in a region of order $\alpha(L)$ around $L$ it is justified to replace $\alpha(L)$ by $\alpha(k)$ in (33)

$$E(k, L) \simeq \frac{1}{\sqrt{2\pi\alpha(k)}}\exp\left(-\frac{(L-k)^2}{2\alpha(k)}\right) \ , \qquad (35)$$

i.e., a Gaussian behaviour for $E(k, L)$ considered as a function of $L$ with $k$ fixed, which is a continuous quantity.

This replacement is justified for GOE and GUE where one has for $L \gg 1$

$$\Sigma^2_{\text{GOE}}(L) \simeq \frac{2}{\pi^2}\left[\ln(2\pi L) + \gamma + 1 - \frac{\pi^2}{8}\right]$$

$$\Sigma^2_{\text{GUE}}(L) \simeq \frac{1}{\pi^2}\left[\ln(2\pi L) + \gamma + 1\right] \ ,$$

with $\gamma = 0.5772\ldots$ being Euler's constant. Because of this logarithmic behaviour the $E_\beta(k, L)$ functions can be considered as Gaussians in $L$.

In the case of classically chaotic systems the $E(k, L)$ functions saturate [31, 42] for some $k \geq k_s$ with $k_s$ being a system and energy dependent constant proportional to $L_{\max}$. Saturation in the $E(k, L)$ functions means that their width does not further increase but rather oscillates around a plateau value. Furthermore, the $E(k, L)$ functions can be well approximated by the Gaussians (35) for $k \geq \min(k_s, k_{\mathrm{RMT}})$. A higher amplitude of the $E(k, L)$ functions corresponds to a smaller width $\alpha(k)$. The saturation of the width $\alpha(k)$ around a plateau $\alpha_\infty$ corresponds to the well-known saturation of the number variance

$$\alpha_\infty = \Sigma_\infty^2 \quad, \tag{36}$$

which again follows from relation (34).

It is worthwhile to note that in RMT $\Sigma^2(L)$ and $\Delta_3(L)$ are related by

$$\Delta_3^{\mathrm{RMT}}(L) = \frac{2}{L^4} \int_0^L \left( L^3 - 2L^2 r + r^3 \right) \Sigma_{\mathrm{RMT}}^2(r) \, dr \quad. \tag{37}$$

Assuming this relation to hold also for an individual system, one can derive [31] that the saturation value $\Sigma_\infty^2$ of the number variance is related to the saturation value $\Delta_\infty$ of the spectral rigidity by

$$\Sigma_\infty^2 \simeq 2\Delta_\infty \quad. \tag{38}$$

Numerical results [11, 28, 31] for strongly chaotic systems suggest that this relation is indeed a good approximation.

After having discussed the consequences of a Gaussian behaviour for the $E(k, L)$ functions for the number variance $\Sigma^2(L)$, we would like to turn to the spectral form factor $F(\tau)$, which in RMT is connected with the $E(k, L)$ functions by the relation [29]

$$F(\tau) = 2\pi\tau \int_0^\infty dL \, \sin(2\pi\tau L) \frac{d}{dL} \Sigma^2(L) \quad. \tag{39}$$

Inserting the approximation (35) in (39) one approximately obtains for $L \gg k_0$, i.e., for $\tau \to 0$,

$$F(\tau) \simeq -4\pi^2\tau^2 \sum_{k=0}^{\infty} \alpha(k) \left( 1 - 4\pi^2\tau^2\alpha(k) \right) e^{-2\pi^2\tau^2\alpha(k)} \cos(2\pi k\tau) \quad, \tag{40}$$

which reproduces the small $\tau$ behaviour of $F(\tau)$ in the case of Poisson and RMT spectra.

# IV    The distribution of $\eta_L(x)$

Before we discuss the distribution of the mode fluctuations $W(x)$, we turn to the distribution of

$$\eta_L(x) := \frac{n(L, x) - L}{\sqrt{\Sigma^2(L, x)}} \quad, \tag{41}$$

where $x$ is randomly chosen from the interval $[T, cT]$ with $c > 1$, and $L > 0$ being a fixed parameter. The distribution of $\eta_L(x)$ with $x, L \to \infty$ is in a certain limit related to $W(x)$. Because of the expectation value $\langle n(L, x) \rangle \simeq L$ and the definition of $\Sigma^2(L, x)$, the distribution of $\eta_L(x)$ for $T \to \infty$ has zero mean and unit variance. In the case of a numerical analysis with a limited number of quantal levels the distribution of $\eta_L(x)$ is not smooth and displays pronounced equidistant peaks since for fixed $L$ the quantity $n(L, x) - L$ is an integer minus a fixed number, and $\Sigma^2(L, x)$ is only slowly varying. A smooth distribution can only be expected in the limit $L, x \to \infty$ because of the increase of $\Sigma^2(L, x)$. There are rigorous results for the distribution of $\eta_L(x)$ in the case of RMT [43] and for some integrable systems [3], see below.

Let us briefly recall some facts about the characteristic function, moments and cumulants of a distribution (see e.g. [44, 45]). The generating function $F(t)$ of a distribution function $P(x)$ is defined by

$$F(t) := \int_{-\infty}^{\infty} t^x \, dP(x) \; . \tag{42}$$

For the special choice $t := e^{iu}$, $u$ real, the characteristic function $f(u)$ of the distribution is obtained

$$f(u) := F\left(e^{iu}\right) = \int_{-\infty}^{\infty} e^{iux} \, dP(x) \; , \tag{43}$$

which exists, since

$$\left| \int_{-\infty}^{\infty} e^{iux} \, dP(x) \right| \leq \int_{-\infty}^{\infty} dP(x) = m_0 = 1 \; .$$

The characteristic function $f(u)$ determines uniquely the distribution $P(x)$. By setting $t := e^\mu$, $\mu$ real, in eq.(42) the moment generating function $M(\mu)$ is obtained

$$M(\mu) := F\left(e^\mu\right) = \int_{-\infty}^{\infty} e^{\mu x} \, dP(x) = \sum_{k=0}^{\infty} m_k \frac{\mu^k}{k!} \; , \tag{44}$$

if the integral exists. Notice, that the convergence is not ensured in contrast to the characteristic function (43). Here $m_k := \int_{-\infty}^{\infty} x^k \, dP(x)$ are the moments of $P(x)$.

The moments $m_k$ of the distribution of $n(L, x)$ can be determined from the $E(k, L)$ functions by the moment generating function

$$M(\mu) = \sum_{k=0}^{\infty} E(k, L) \, e^{k\mu} = \sum_{k=0}^{\infty} m_k \frac{\mu^k}{k!} \; , \tag{45}$$

whenever the series are convergent. For a centered Gaussian with variance $\sigma^2$ the odd moment vanish, and the even moments are given by

$$m_{2k} = \frac{(2\sigma^2)^k}{\sqrt{\pi}} \, \Gamma(k + 1/2) = \frac{(2k)!}{k! \, 2^k} \sigma^{2k} \; . \tag{46}$$

The even moments are increasing with increasing $k$. If one deals with a Gaussian distribution, it is thus more desirable to use the cumulants instead of the moments, since the cumulants $c_k$ have the advantage that for a Gaussian distribution only the first two cumulants are non-vanishing. The cumulants $c_k$, $k \geq 2$, have the additional advantage that they are translational invariant under a shift in the variable $L$ in contrast to the moments, i.e., the origin, with respect to which they are measured, plays no role, except for $c_1 = m_1$. Thus, the cumulants $c_k$, $k \geq 2$, for $n(L, x)$ and for $n(L, x) - L$ are identical. If one transforms a distribution into the so–called *standard measure*, i.e., with zero mean and unit variance, one has

$$C_1 := 0 \quad \text{and} \quad C_k := \frac{c_k}{c_2^{k/2}} \; . \tag{47}$$

The standard measure of $n(L, x)$ is $\eta_L(x)$, of course. To obtain the cumulants $C_k$ of $\eta_L(x)$, we first compute the cumulants $c_k$ of $n(L, x)$, and then transform them according to (47). If the cumulants $C_k$ with $k \geq 3$ vanish for $L \to \infty$, the distribution of $\eta_L(x)$ is a Gaussian in the weak limit. The cumulants have, however, the disadvantage that they cannot be computed directly from the data, i.e., with integrative or summatory processes. They can be computed from the moments or from the cumulant generating function, respectively, which is related to the moment generating function. The cumulant generating function reads

$$C(\mu) = \ln M(\mu) = \sum_{k=1}^{\infty} c_k \frac{\mu^k}{k!} \quad , \tag{48}$$

so that the cumulants are given by

$$c_k = \left. \frac{d^k}{d\mu^k} C(\mu) \right|_{\mu=0} \quad . \tag{49}$$

In RMT the asymptotic behaviour for $L \to \infty$ of the cumulants $c_k$ for the distribution of $n(L, x)$ is studied in [43]. Because of the stationarity of the random matrix ensembles one sets the energy variable $x = -L/2$ and considers the probabilities with respect to the interval $[-L/2, L/2]$. The second cumulant is $c_2(L) = \frac{1}{\pi^2} \ln L + O(1)$ for the GUE case and, therefore, the $k$th cumulant of $\eta_L$ will be $C_k \simeq c_k / (\frac{1}{\pi^2} \ln L)^{k/2}$. Thus a Gaussian behaviour is displayed if $C_k \to 0$ for $k \geq 3$, i.e., $c_k = o\left((\frac{1}{\pi^2} \ln L)^{k/2}\right)$ for $k \geq 3$. This behaviour is shown to be true for GUE, and a similar behaviour is obtained for GOE and GSE in [43]. Therefore *the distribution of $\eta_L$ is Gaussian in at least these three RMT ensembles in the weak limit for $L \to \infty$.*

For integrable systems with eigenvalues $E_{n_1, n_2} = I(n_1 - \alpha_1, n_2 - \alpha_2)$, where $I(x_1, x_2)$ is a smooth homogeneous function of second degree and $(\alpha_1, \alpha_2)$ is a point in the unit sphere, the distribution of $n(L, x)$ is studied in [3, 46]. It is proven that $N(x + L)$ and $N(x)$ are asymptotically independent for $L/x \to 0$ and $L/\sqrt{x} \to \infty$. This limit can be realized by the choice $x \to \infty$ with $L = zx^\delta$, $\frac{1}{2} < \delta < 1$.

Furthermore $\eta_L(x)$ converges to a difference of two independent identically distributed random variables, whose limit distributions are the same as for $W(x)$. In the limit $x \to \infty$ with $L/\sqrt{x} \to z > 0$ it is proven that $\eta_L(x)$ has a non-Gaussian limit distribution. However, for $z \to 0$ this distribution tends to a normal distribution. In addition rigorous results concerning the behaviour and saturation of the number variance can be found in [3, 46, 47].

For the chaotic quantum systems discussed later, a Gaussian behaviour is observed for the $E(k, L)$ functions for sufficiently large $k$ values as a function of $L$, see eq.(35). Due to the saturation this implies that the approximation (33) is fulfilled for large $L$. Using this approximation in (45) and (48) one obtains a series, which is approximated by an integral using the Euler-MacLaurin summation formula. Evaluating the integral yields for large $L$

$$C(\mu) \simeq L\mu + \frac{\alpha(L)}{2} \mu^2 \quad , \quad L \gg 1 \quad . \tag{50}$$

Therefore one obtains $c_1 \to L$, $c_2 \to \alpha(L)$ and $c_k \to 0$ for $k \geq 3$ which yields $C_k = 0$, $k \geq 3$, of course. Thus *the distribution of $\eta_L = (n(L) - L)/\sqrt{\alpha(L)}$ is a normalized Gaussian in the weak limit $L \to \infty$, if the Gaussian behaviour is established for $E(k, L)$.*

In the case of a Poissonian random process one obtains with formula (16) for $E_{\text{Poisson}}(k, L)$, using (45) and (48), the cumulant generating function

$$C(\mu) = L(e^\mu - 1) \quad . \tag{51}$$

Together with eq.(49) it immediately follows that for all cumulants $c_k = L$ in the case of a Poissonian random process. Since $C_k = L/L^{k/2} \to 0$, $k \geq 3$, for $L \to \infty$, the *distribution of $\eta_L$ approaches a Gaussian for a Poissonian random process in the sense of a weak limit.*

# V  Mode-fluctuation distribution $P(W)$

As already discussed, the description of spectral statistics using finitely many energy levels by means of RMT is restricted to short-range correlations because of, e.g., the saturation of the number variance and rigidity, contrasting the expected logarithmic increase in RMT. Furthermore, even for short-range correlations deviations were found for a class of strongly chaotic systems, the so-called arithmetical systems [9, 10, 11, 13]. Therefore in [14, 15] a different statistical measure as signature of quantum chaos is proposed which is given by the probability distribution of the suitably normalized fluctuations of $N_{\text{fluc}}(x)$.

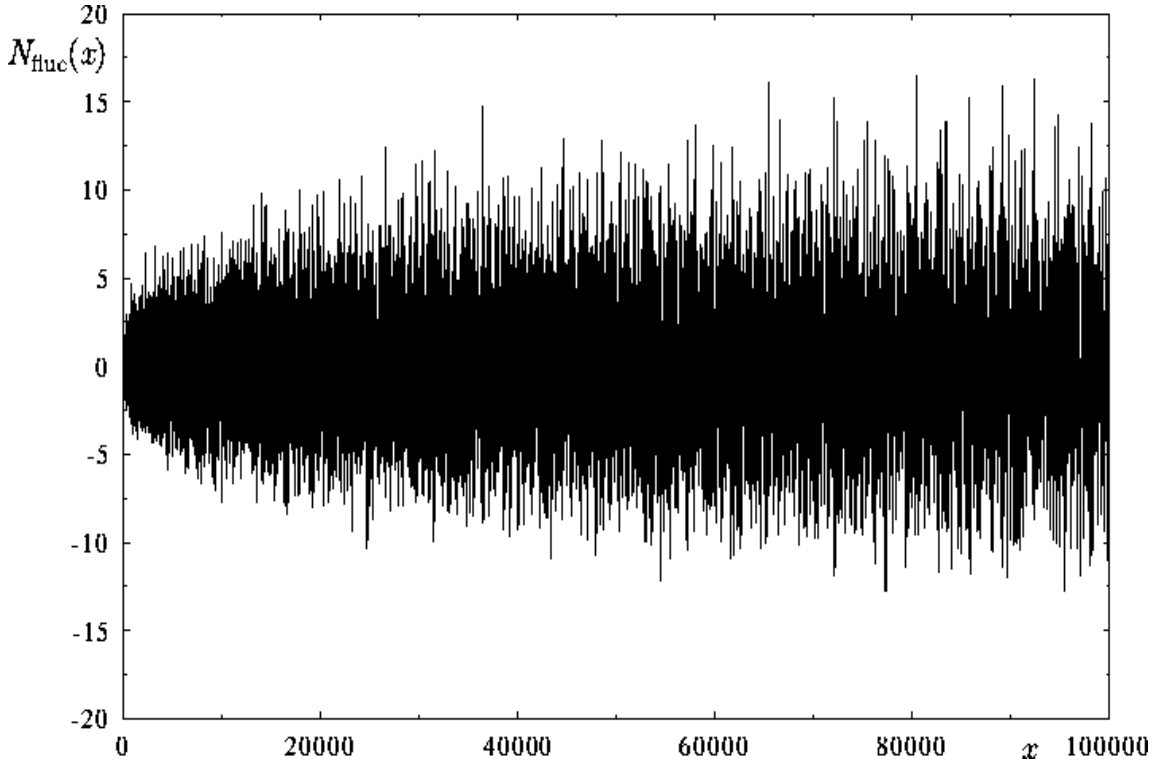

Figure 3: $N_{\text{fluc}}(x)$ for the integrable circular billiard.

In order to define this statistical quantity, some considerations in advance are necessary. $N_{\text{fluc}}(x)$ is shown in fig. 3 for the integrable circular billiard (see sec. VII.2), using the first 100 000 eigenvalues of the even symmetry class with respect to reflection at the diameter. Since $N_{\text{fluc}}(x)$ behaves very irregularly one might consider $N_{\text{fluc}}(x)$ as a random function of $x$ and look for its statistical properties. The increase in the amplitude of the fluctuations is clearly visible in fig. 3. Therefore, one defines the variance

$$D(T) := \frac{\Xi(c)}{(c-1)T} \int_T^{cT} [N_{\text{fluc}}(y)]^2 \, dy \quad , \qquad (52)$$

with $c > 1$, where $\Xi(c)$ is a correction to be explained below. With this one can construct the following random variable

$$W(x) := \frac{N_{\text{fluc}}(x)}{\sqrt{D(x)}} \quad , \qquad (53)$$

where $x$ is randomly chosen from the interval $[T, cT]$. This statistic was first investigated in the special case of the eigenvalues of the Laplacian on a torus [27]. A similar statistic for the GUE without the normalization was studied in [48]. If a limit distribution for $T \to \infty$ of this random variable $W(x)$ exists, it has by construction a second moment of 1 if the second moment exists. Furthermore, the first moment of this limit distribution is zero because by definition $N_{\text{fluc}}(x)$ describes the fluctuations of $N(x)$ around the mean behaviour $\overline{N}(x) = x$.

The conjecture proposed in [15] can now be formulated as follows:

*For bound conservative and scaling systems the quantity $W(x)$, eq. (53), possesses a limit distribution for $T \to \infty$. This distribution is absolutely continuous with respect to the Lebesgue measure on the real line, with a density $P(W)$*

$$\lim_{T \to \infty} \frac{1}{(c-1)T} \int_T^{cT} g(W(x))\, \rho(x/T)\, dx = \int_{-\infty}^{\infty} g(W)\, P(W)\, dW \ , \tag{54}$$

*where $g(x)$ is a bounded continuous function, and $\rho(t) \geq 0$ is a continuous density on $[1, c]$ with $\frac{1}{c-1} \int_1^c \rho(t)\, dt = 1$.*

*Furthermore, the limit distribution has zero mean and unit variance,*

$$\int_{-\infty}^{\infty} dW\, W\, P(W) = 0 \quad , \quad \int_{-\infty}^{\infty} dW\, W^2\, P(W) = 1 \ . \tag{55}$$

*If the corresponding classical system is strongly chaotic, having only isolated and unstable periodic orbits, then $P(W)$ is universally a Gaussian, $P(W) = \frac{1}{\sqrt{2\pi}} \exp(-\frac{1}{2} W^2)$. In contrast, a classically integrable system leads to a non-Gaussian density $P(W)$.*

In the case of several integrable systems eq. (54) is proven, and furthermore, it is shown that the corresponding limit distribution decays faster than a Gaussian. These results will be summarized in the following section.

Some comments are in order:

First notice that eqs. (55) do not follow from (54) by choosing $g(x) = x$ and $g(x) = x^2$, respectively, since $g$ is supposed to be bounded. The question of the existence of higher moments is rather non-trivial, as will become clear in the case of integrable systems in the following section.

Secondly, if a chaotic system has non-isolated periodic orbits, the conjecture can be generalized by taking their contribution $N_{\text{long}}(E)$ into account in "Weyl's law" $\overline{N}(E)$, eq. (4). Unfolding the energy levels in this way yields a Gaussian behaviour for the distribution of $W(x)$. Notice that in the case of other long-range contributions which vanish for $x \to \infty$, it is sometimes necessary to take them into account in $N_{\text{long}}(x)$ in order to approximate the limit distribution sufficiently well with only a finite number of energy levels.

Thirdly, note that the limit distribution $P(W)$ is independent of the choice of the probability density $\rho(t)$. This is easily established, if one can prove eq. (54) in the case of $\rho(t) = 1$, by approximating an arbitrary $\rho(t) \in L^1([1, c])$ by step-wise functions (see [49, 50]). By choosing different densities $\rho(t)$ the distribution of $x$ on $[T, cT]$ is $\rho(x/T)\, dx/T$, which enables one to make a change of variables [3, 47]. E.g., one can choose $\rho(t) = 1$, which corresponds to an equidistribution of $x$ on $[T, cT]$. Choosing $\rho(t) = \frac{c-1}{2(\sqrt{c}-1)} t^{-1/2}$ corresponds to averaging with respect to $d(x^{1/2})$, i.e., $\sqrt{x}$ is distributed uniformly on $[\sqrt{T}, \sqrt{cT}]$. Similarly for $\rho(t) = \frac{2}{c+1} t$ one obtains an equidistribution of $x^2$ on $[T^2, (cT)^2]$.

Berry's semiclassical analysis [2] provides predictions of the asymptotic behaviour $x \to \infty$ of the spectral rigidity $\Delta_3(L, x)$, i.e., about the energy dependence of the saturation value $\Delta_\infty(x)$ for $x \to \infty$. Analogous arguments carry over to the energy dependence of the variance $D(x)$. For generic integrable billiards one obtains

$$D(x) \simeq c_{\text{Int}} \sqrt{x} \; , \tag{56}$$

where $c_{\text{Int}}$ is some non-universal constant which depends on the considered system as well as on the chosen average. In the case of generic classically chaotic systems with anti-unitary symmetry (e. g. time-reversal symmetry), these arguments yield

$$D(x) \simeq \frac{1}{2\pi^2} \ln x + C \; , \tag{57}$$

and for systems without such a symmetry

$$D(x) \simeq \frac{1}{4\pi^2} \ln x + C' \; . \tag{58}$$

In the case of arithmetical chaotic systems arguments based on the diagonal approximation of the spectral form factor lead to [51]

$$D(x) \simeq \frac{\sqrt{\mathcal{A}}}{\pi^{3/2}} \frac{\sqrt{x}}{\ln x} \; , \tag{59}$$

where $\mathcal{A}$ is the area of the corresponding arithmetic surface. However, the arithmetic hyperbolic triangle discussed in section VII.1 possesses a behaviour which is more conform to $c_1\sqrt{x} + c_2$ like integrable systems [11]. These results can be used in asymptotic considerations as analytic expressions for the variance $D(x)$ (see section VII.3 for details concerning numerical tests of the conjecture).

We are now in a position to explain the factor $\Xi(c)$ occurring in the variance $D(x)$ in eq.(52). To obtain a distribution of $W(x)$ with variance one it is necessary to normalize $N_{\text{fluc}}(x)$ with the variance at $x$. However, with $\widehat{D}(x) := \langle [N_{\text{fluc}}(x')]^2 \rangle_x$ one uses a variance corresponding to a point $x' \in [x, cx]$ for the normalization which differs from the local variance at $x$ for large averaging parameter $c$ or for sufficiently increasing $D(x)$. Therefore we now study the alteration caused by the averaging (8) in the functional dependence on $D(x)$.

Considering a power law for $D(x) = ax^\gamma$ which includes (56) with $\gamma = \frac{1}{2}$, one obtains

$$\widehat{D}(x) = \langle D(x') \rangle_{\hat{x}} = \frac{c^{\gamma+1} - 1}{(\gamma + 1)(c - 1)} D(x) \tag{60}$$

which in turn leads to

$$\Xi(c) = \frac{3(c-1)}{2(c^{3/2} - 1)} \quad \text{for} \quad \gamma = \frac{1}{2} \; . \tag{61}$$

For a functional dependence $D(x) = a \ln x + b$, covering the chaotic cases (57) and (58), one obtains

$$\widehat{D}(x) = D(x) + a \left( \frac{c \ln c}{c - 1} - 1 \right) \; . \tag{62}$$

Thus for chaotic systems (excluding arithmetical chaos) one gets $\Xi(c) = 1$. Asymptotically the constant plays no role, but for numerical computations it is sometimes necessary to take this additive constant into account.

The above formulation of the conjecture is slightly different from the one given in [15]. Therefore let us now describe in short the connection between both formulations: In [15] the mode-fluctuating distribution was introduced for bound conservative scaling systems with at least two degrees of freedom. There the variable $\tilde{x} = E^\alpha$, with $\alpha$ being the scaling parameter, was used as the basic variable. From a theoretical point of view this choice seems more appropriate because for scaling systems the natural variable is $\tilde{x}$ in Gutzwiller's trace formula [7]. For numerical considerations, however, it is much better to have an equidistribution on the whole unfolded energy interval, which gives every level the same "weight". For example, a uniformly distributed random variable in $p$ pronounces the contribution of the lower energy levels, which is not desired. Nevertheless, there is no contradiction between both formulations, since both are equivalent, which can be shown by appropriate choices of the density $\rho(t)$:

Choosing $\rho(t) = \frac{\alpha(c-1)}{c^\alpha - 1} t^{\alpha - 1}$, one can change from the $\tilde{x}$ formulation to the formulation in the energy variable $E$. The reverse direction is obtained by $\rho(t) = \frac{c-1}{(c^{1/\alpha}-1)\alpha} t^{\frac{1}{\alpha} - 1}$. If one neglects for large energies $E$ the circumference and the constant term in the generalized Weyl's law (4) it is trivial that the formulation in $E$ and the one in the unfolded variable $x$ are equivalent.

For classically chaotic systems a semiclassical approximation for $N_{\text{fluc}}(x)$ is given by means of Gutzwiller's periodic-orbit sum. Here $N_{\text{fluc}}(x)$ can be expressed using the *dynamical zeta function $Z(s)$* which involves only properties of periodic orbits like their lengths, stabilities and Maslov indices. The non-trivial zeros of the dynamical zeta function $Z(s)$ are directly connected with the quantal levels of the corresponding system. Using the argument principle, one obtains

$$\widetilde{N}_{\text{fluc}}(p) = \frac{1}{\pi} \arg Z(ip) \ . \tag{63}$$

The tilde is chosen to distinguish the different arguments, i.e., the momentum $p$ instead of the unfolded energy $x$.

An interesting example is provided by the non-trivial zeros of the Riemann zeta function $\zeta(s)$. Their statistics behave in many respects like the eigenvalues of some hypothetical classically chaotic system without anti-unitary symmetry. Here one has

$$\widetilde{N}_{\text{fluc}}(p) = \frac{1}{\pi} \arg \zeta\left(\frac{1}{2} + ip\right) \ .$$

It can be shown that for a certain class of $\zeta$-functions, which includes the Riemann zeta function, the corresponding $\widetilde{W}(p) = \frac{\widetilde{N}_{\text{fluc}}(p)}{\sqrt{\ln \ln p}}$ has a Gaussian limit distribution [52, 53], see also [54].

In order to shed some more light on the meaning of the conjecture, we emphasize the fact that a Gaussian limit distribution reflects the validity of a *central limit theorem* for the fluctuations. If one considers all probability distributions with density and with unit variance, then the *spectral entropy*

$$\mathcal{E}[P] := - \int\limits_{-\infty}^{+\infty} dW \ P(W) \ln P(W)$$

is maximized for a Gaussian distribution with mean zero. Since $\mathcal{E}[P]$ measures the mean unlikelihood for $W(x)$ to be of a specific value, it provides a quantitative measure for spectral randomness. Thus the content of the conjecture is that *classically strongly chaotic systems have maximally random quantum spectra*. We should mention that from the numerical point of view $\mathcal{E}[P]$ is not a suitable quantity to distinguish between a Gaussian and a different limit distribution [42].

Before we present the relation between $\eta_L(x)$ and $W(x)$, we need a relation between the number variance $\Sigma^2(L, \hat{x})$ and $D(\hat{x})$. To be consistent with the definition of the variance $D(\hat{x})$,

eq.(52), we use an analogous definition for $\Sigma^2(L,\hat{x})$, which gives the number variance at $\hat{x}$ and not at some $x \in [\hat{x}, c\hat{x}]$ as explained above. Such a definition differs from (7) only for systems with a non-logarithmic behaviour of $D(\hat{x})$. Thus we get

$$\begin{aligned}
\Sigma^2(L,\hat{x}) &= \frac{\Xi(c)}{(c-1)\hat{x}} \int_{\hat{x}}^{c\hat{x}} dx \, [N(x+L) - N(x) - L]^2 \\
&= \frac{\Xi(c)}{(c-1)\hat{x}} \int_{\hat{x}}^{c\hat{x}} dx \, [N_{\text{fluc}}(x+L) - N_{\text{fluc}}(x)]^2 \\
&= \frac{\Xi(c)}{(c-1)\hat{x}} \int_{\hat{x}}^{c\hat{x}} dx \, \left([N_{\text{fluc}}(x+L)]^2 - 2N_{\text{fluc}}(x+L)N_{\text{fluc}}(x) + [N_{\text{fluc}}(x)]^2\right) \ .
\end{aligned}$$

If one considers the limit of large $\hat{x}$ and $L$ in such way that one is in the saturation regime, i.e., $L \gg L_{\max} = C\sqrt{\hat{x}}$, one expects $N_{\text{fluc}}(x+L)$ and $N_{\text{fluc}}(x)$ to become independent, which is proven in [3] for a class of integrable systems. Under this assumption the integral over $N_{\text{fluc}}(x+L)N_{\text{fluc}}(x)$ vanishes. The last summand yields $D(\hat{x})$, and one can show that the first summand yields $D(\hat{x})$, too. To that end let us fix $L = C\hat{x}^\delta$, with $\frac{1}{2} < \delta < 1$, and consider the first term for $x \to \infty$

$$\int_{\hat{x}}^{c\hat{x}} dx \, [N_{\text{fluc}}(x+L)]^2 = \int_{\hat{x}(1+C\hat{x}^{\delta-1})}^{c\hat{x}(1+\frac{C}{c}\hat{x}^{\delta-1})} dx \, [N_{\text{fluc}}(x)]^2 \to \int_{\hat{x}}^{c\hat{x}} dx \, [N_{\text{fluc}}(x)]^2 = \frac{(c-1)\hat{x}}{\Xi(c)} D(\hat{x}) \ .$$

For large $\hat{x}$ the upper bound tends to $c\hat{x}$ and the lower bound to $\hat{x}$, because of $\delta < 1$. Thus we finally have

$$\Sigma^2(L,\hat{x}) \simeq 2\,D(\hat{x}) \qquad \text{for large } \hat{x} \text{ and } L = C\hat{x}^\delta, \ \tfrac{1}{2} < \delta < 1 \ . \qquad (64)$$

In comparison with $\eta_L(x)$, the $W(x)$ statistic has the advantage that one need not care about the different cases of $\hat{x}, L \to \infty$, because one is always in the non-universality regime $L \gg L_{\max}$. Both statistics are related (the basic idea is due to Bolte [55]) in the following way: If one considers in

$$\eta_L(x) = \frac{N(x+L) - N(x) - L}{\sqrt{\Sigma^2(L,x)}} = \frac{[N(x+L) - (x+L)] - [N(x) - x]}{\sqrt{\Sigma^2(L,x)}} \qquad (65)$$

the limit of large $x$ in such a way that one is in the saturation regime by setting as above $L = Cx^\delta$, $\frac{1}{2} < \delta < 1$, we can use (64)

$$\eta_L(x) \simeq \frac{1}{\sqrt{2}} \left[ \frac{[N(x+L) - (x+L)]}{\sqrt{D(x)}} - \frac{[N(x) - x]}{\sqrt{D(x)}} \right] \ . \qquad (66)$$

The second fraction in the bracket is $W(x)$. If in the limit of large $x, L$ with $L = Cx^\delta$, $\frac{1}{2} < \delta < 1$, the random variables $N(x+L)$ and $N(x)$ are statistically independent, the right hand side is the sum of two independent random variables. Assuming that $\eta_L(x)$ is Gaussian distributed for $L \to \infty$, then each of the random variables on the r.h.s. has a Gaussian distribution [44].

Furthermore, if $W(x)$ has a Gaussian distribution, also the first term in eq.(66) has a Gaussian distribution. If both random variables are independent, $\eta_L(x)$ is also Gaussian distributed.

The variance is one, because the variance of the sum of two independent random variables is the sum of each variance [44].

Thus for $x, L \to \infty$ with $L = Cx^\delta$, $\frac{1}{2} < \delta < 1$, a Gaussian behaviour of $\eta_L(x)$ implies a Gaussian behaviour of $W(x)$ and vice versa.

Assuming the conjecture, it follows that in the above limit the energy levels of integrable systems cannot behave like the outcomes of a Poissonian random process, since a Poissonian distribution for the $E(k, L)$ functions implies a Gaussian distribution of $\eta_L$, which has a Gaussian distribution for $W(x)$ as a consequence.

Furthermore, it should be remarked that the above discussion shows that $W(x)$ is a statistic which measures fluctuations in the saturation regime, i.e., global fluctuations.

## VI  Exact results for integrable systems

Recently a lot of results concerning the limit distribution $P(W)$ of $W(x)$ in the case of integrable systems were obtained (summaries can be found in [46, 56, 57]). For several integrable systems (see below) it is possible to show that $N(E)$ can be decomposed as follows

$$N(E) = \frac{\mathcal{A}}{4\pi} E + E^{1/4} \, \Theta(\sqrt{E}) \ .$$

The remainder term $E^{1/4} \, \Theta(\sqrt{E})$ fluctuates around zero with an amplitude increasing as $E^{1/4}$ and corresponds to $N_{\text{fluc}}(E)$. The first term is Weyl's law. Since the considered systems have no boundary, there is no additional term involving $\mathcal{L}$, and also the constant $\mathcal{C}$ is zero.

The basic idea for proving the existence of a limit distribution of $\Theta(\sqrt{E})$ is to express $N(E)$ in terms of a lattice-point problem and then to exploit the Poisson summation formula to obtain an expression of the remainder term as a trigonometric series. If the remainder term is an *almost periodic function* [58] (see also [49, 56]) then it has a limit distribution [49, 59]. To be more precise, we have to give some definitions (following the exposition in appendix II of [56]):

The space $B^p$ of $p$th *power summable Besicovitch almost periodic functions* is defined as follows: $f \in B^p$, if for any $\epsilon > 0$ there exists a finite set of numbers $(a_k, \omega_k)$, $k = 1, \ldots, N(\epsilon)$, $\omega_k \in \mathbf{R}$, $a_k \in \mathbf{C}$, such that

$$\limsup_{T \to \infty} \left[ \frac{1}{T} \int_0^T \left| f(t) - \sum_{k=1}^{N(\epsilon)} a_k \exp(i\omega_k t) \right|^p dt \right]^{1/p} < \epsilon \ .$$

This means that every $f \in B^p$ can be approximated in the mean arbitrarily well by trigonometric series. In $B^p$ one has the semi-norm

$$||f||_{B^p} := \limsup_{T \to \infty} \left[ \frac{1}{T} \int_0^T |f(x)|^p \, dx \right]^{1/p} ,$$

such that $||f - g||_{B^p} = 0$ for $f, g \in B^p$ defines an equivalence relation. For $1 \le p_1 < p_2$ one has $B^{p_2} \subset B^{p_1}$ and for $f \in B^{p_2}$ the inequality $||f||_{B^{p_2}} \le ||f||_{B^{p_1}}$ holds.

A formal trigonometric series $\sum_{k=1}^{\infty} a_k \exp(-i\omega_k t)$ is called generalized Fourier series of $f$ with respect to $B^p$, if $\lim_{N \to \infty} ||f(t) - \sum_{k=1}^{N} a_k \exp(-i\omega_k t)||_{B^p} = 0$. For a given $f \in B^p$ its generalized Fourier series is well defined and unique.

Let us further define the space $H$, with $f \in H$, if for any $\epsilon > 0$, there exists a finite set of numbers $(a_k, \omega_k)$, $k = 1, \ldots, N(\epsilon)$, $\omega_k \in \mathbf{R}$, $a_k \in \mathbf{C}$, such that

$$\limsup_{T \to \infty} \frac{1}{T} \int_0^T \min\left(1, \left| f(t) - \sum_{k=1}^{N(\epsilon)} a_k \exp(i\omega_k t) \right| \right) dt < \epsilon \; .$$

This assumption is weaker than $f \in B^1$, since $B^1 \subset H$.

In [49] it is proven using ergodic-theoretic and probabilistic methods that the condition $f \in H$ is sufficient to establish that $f(t)$ possesses a limit distribution

$$\lim_{T \to \infty} \frac{1}{T} \int_0^T g(f(t)) \, dt = \int_{-\infty}^{+\infty} g(x) \nu(dx)$$

for every bounded continuous function $g(x)$ on $\mathbf{R}$. If in addition $f \in B^p$, $p \geq 1$, then

$$||f||_p = \left[ \int_{-\infty}^{+\infty} |x|^p \, \nu(dx) \right]^{1/p} < \infty \; , \qquad \lim_{T \to \infty} \frac{1}{T} \int_0^T f(t) dt = \int_{-\infty}^{+\infty} x \, \nu(dx) \; .$$

Thus for $f \in B^p$ the absolute moments up to the $p$th order exist.

In a series of papers it is proven for several integrable systems that the corresponding normalized remainder term $\Theta(\sqrt{E})$ belongs to $H$, or even $B^p$. In these proofs $N(E)$ is expressed in terms of a lattice-point problem. A very old problem of this type, due to Gauß, is the classical circle problem, where one considers the difference between the number of lattice points inside a circle of radius $R$ and the corresponding area $\pi R^2$. For a long time the aim was to find uniform bounds of the remainder term. In [59] the study of limit distributions of the remainder term by means of almost periodic functions was started. From our point of view one is led to this problem if one considers the eigenvalues of the Laplace operator on a two-dimensional torus. These read (in suitable units) $E_{n,m} = n^2 + m^2$, $n, m \in \mathbf{Z}$. Thus $N(E)$ is given by the number of lattice points inside a circle of radius $R = \sqrt{E}$. Applying the Poisson summation formula gives

$$N_{\text{fluc}}(R^2) = N(R^2) - \pi R^2 = \sum_{n \in \mathbf{Z}^2 \setminus \{(0,0)\}} \frac{J_1(2\pi |n| R)}{|n|} R \; ,$$

where $|n| = \sqrt{n_1^2 + n_2^2}$. Using the asymptotic behaviour of the Bessel function one obtains for the normalized mode fluctuations $\Theta(R) = N_{\text{fluc}}(R^2)/\sqrt{R}$

$$\Theta(R) = \frac{1}{\pi} \sum_{n \in \mathbf{Z}^2 \setminus \{(0,0)\}} \frac{\cos(2\pi |n| R - \frac{3\pi}{4})}{|n|^{3/2}} + O(R^{-1}) \; .$$

The normalized remainder term can then be rewritten as a sum over square free numbers $m$

$$\Theta(R) = \sum_{\text{square free } m} f_m(\sqrt{m} R) + O(R^{-1}) \; ,$$

with

$$f_m(t) = \frac{1}{\pi m^{3/4}} \sum_{k=1}^{\infty} \frac{g_{k,m}}{k^{3/2}} \cos(2\pi k t - \frac{3\pi}{4}) \; ,$$

where $g_{k,m} := \#\{n \in \mathbf{Z}^2 \mid |n| = k\sqrt{m}\}$. The above representation of $\Theta(R)$ can be interpreted as a flow on an infinite-dimensional torus with incommensurate frequencies, where $R$ is considered as the time-parameter. By this the problem of finding a limit distribution becomes a problem in ergodic theory. The above representation can also be interpreted as a trace formula, where the sum runs over the lengths $\sqrt{n_1^2 + n_2^2}$ of primitive periodic orbits of the geodesic motion on a two-dimensional torus, and the sum in the representation of $f_m(t)$ corresponds to multiple traversals. It is proven in [59] that the corresponding normalized remainder term $\Theta(\sqrt{E})$ belongs to $B^2$ and possesses a limit distribution. Furthermore, the limit distribution is absolutely continuous with respect to the Lebesgue measure, i.e., it can be written as $\nu(dx) = P(x)\,dx$ by means of a density $P(x)$. In addition, estimates for derivatives of the density are given and the convergence of the (absolute) moments is proven up to the ninth moment. In [60] it is proven that the third moment is negative, which implies that the limit distribution is non-Gaussian.

In [50] the more general case of the number of lattice points inside a shifted circle is considered. It is proven that the corresponding normalized remainder term has an absolutely continuous limit distribution and that the first and the second moments exist. The corresponding probability density $P(x)$ decays faster than a Gaussian: for every $\epsilon > 0$ and $|x| > x_0(\epsilon)$

$$0 \leq P(x) < \exp(-|x|^{4-\epsilon}) \;, \tag{67}$$

which again shows that the limit distribution cannot be Gaussian. In addition in [61] it is proven that the variance of the remainder term is a wild function of the shift of the circle.

In [49] the existence of a limit distribution and the first two moments is shown for the remainder term $\Theta(\sqrt{E})$ of the number of lattice points inside a smooth strictly convex shifted oval including the origin.

In [62] the existence of a limit distribution for the normalized remainder term of the shifted Laplace operator $(i\vec{\nabla} - \vec{\alpha})^2$ on the two-dimensional torus $\mathbf{T}^2(2\pi a_1, 2\pi a_2)$ is proven. It is shown for transcendental ratios $a_1/a_2$ that the limit distribution possesses a density, which fulfills a slightly stronger condition than (67). In addition these results are extended to a general class of convex ovals. From these results it easily follows that for the integrable rectangular billiard with sides $a$ and $b$ and for the right-angled isosceles triangular billiard the corresponding limit distribution is non-Gaussian.

In [63] the existence of the limit distribution is proven for the eigenvalues of the Laplace operator on certain surfaces of revolution. In particular, in this case $N(E)$ can be written in terms of a lattice-point problem for the Einstein-Brillouin-Keller quantization. It is proven that the normalized remainder term is of Besicovitch class $B^2$ and its generalized Fourier series is given by a sum over closed geodesics. Under an additional assumption it could be shown that the limit distribution has a density, which fulfills $0 \leq P(x) \leq C \exp(-\lambda x^4)$, $C, \lambda > 0$. Also in [63] a more general class of surfaces of revolution is studied, where the remainder term can be decomposed in the following way

$$N_{\text{fluc}}(E) = N(E) - \frac{\mathcal{A}}{4\pi}E \;=\; E^{1/3}\,\Phi(\sqrt{E}) + E^{1/4}\,F(\sqrt{E}) \;. \tag{68}$$

Here $\Phi(\sqrt{E})$ is a finite sum of bounded periodic functions, which implies that its distribution has compact support. $F(\sqrt{E})$ is an almost periodic function of Besicovitch class $B^2$, and thus has a limit distribution. In the above case the leading behaviour of $N_{\text{fluc}}(E)$ is determined by the first term $E^{1/3}\,\Phi(\sqrt{E})$. Considering the behaviour of $F(\sqrt{E})$ corresponds to study the distribution of $[N_{\text{fluc}}(E) - E^{1/3}\,\Phi(\sqrt{E})]/E^{1/4}$. This way of splitting is similar to the one introduced in eq. (4) involving the term $N_{\text{long}}(E)$.

In [56] the remainder term for the eigenvalues of the Laplace-Beltrami operator on a two-dimensional torus with a Liouville metric is investigated. The geodesic motion on such Liouville surfaces can be seen as the most general case which is completely integrable. Using a refined Einstein-Brillouin-Keller quantization the study of the remainder term is reduced to a lattice-point problem. It is proven under a certain condition of non-degeneracy that the normalized remainder term is of Besicovitch class $B^1$. For generic surfaces the distribution of $\Theta(\sqrt{E})$ has a density, which decays faster than a Gaussian. In [64] a refined estimate for the decay of the distribution is proven. Also an explicit expression for the Fourier transform of the density by means of the Fourier amplitudes of $\Theta(\sqrt{E})$ is given. In addition a trace formula for Liouville surfaces is derived.

In [65] the distribution of eigenvalues of Schrödinger operators on manifolds all of whose geodesics are closed are studied. It is proven that the corresponding $\Theta(x)$ is in $B^2$, and that the limit distribution is an equidistribution on a finite interval. As a remark one should add that in the case of the one-dimensional harmonic oscillator and of a particle in a one-dimensional box the corresponding limit distribution is also a box-shaped function.

# VII  Discussion of the considered systems

## VII.1  Models for quantum chaos on surfaces of negative curvature

The motion of a point particle on surfaces of negative curvature is of special historical interest in the context of chaos since already in 1898 with them Hadamard provides the first example of strongly chaotic systems [66]. In 1939 Hopf proves that negative curvature necessarily leads to ergodicity [67], and for the special case of constant negative curvature it is later proven that it leads to strongly mixing systems and its time-evolution operator has Lebesgue spectrum [68]. For compact Riemann surfaces of constant negative curvature Sinai proves that the systems belong to the class of $K$-systems [69], i.e., they show hard chaos. However, the special role these surfaces play is not only of historical nature, since, in addition to the proven property of being a $K$-system, they also possess due to the Selberg trace formula [70] an exact periodic-orbit trace formula. The latter is valid only semiclassically for general systems [7].

A convenient model for surfaces of constant negative curvature is the Poincaré disc $\mathcal{D}$, consisting of the interior of the unit circle in the complex $z$-plane ($z = x_1 + ix_2$), endowed with the metric

$$g_{ij} = \frac{4}{(1 - x_1^2 - x_2^2)^2} \delta_{ij} \quad , \quad i,j = 1,2 \quad . \tag{69}$$

This metric has constant negative Gaussian curvature $K = -1$. This choice fixes the length scale. The classical dynamics is determined by the Hamiltonian

$$H = \frac{1}{2m} p_i\, g^{ij} p_j \quad , \quad p_i = m g_{ij} \frac{dx^j}{dt} \quad . \tag{70}$$

To obtain a compact Riemann surface from the infinitely extended hyperbolic plane one tessellates the plane into copies of one fundamental domain by the action of a Fuchsian group. This can be viewed as cutting out a polygon from the plane in such a way that opposite edges can be glued together, i.e., periodic boundary conditions are imposed [7, 71, 72]. The simplest compact case allowed is a tiling polygon with eight edges, yielding a surface of genus two, which is topologically a sphere with two handles. For more details about these hyperbolic octagons see [7, 72].

The set of all octagons provides an ensemble of $K$-systems, which can be investigated in the sense of RMT, i.e., an ensemble of systems with the same dynamical properties. The usual statistical analysis of a quantum chaotic system concerns in general only an individual system. In the case of the octagon ensemble, however, there is the possibility to study the statistical properties of a physical ensemble, where individual properties are averaged out. A statistical analysis of the $E(k, L)$ distributions, and other important statistics of the octagon ensemble, is carried out in [31]. To our knowledge this is the only case where a statistical analysis of physical systems is carried out in the spirit of RMT. But since the main conjecture of quantum chaos is that the statistical behaviour of RMT is valid for a typical individual quantum chaotic system, at least at short-range correlations, we would like to study the properties of an individual octagon instead of a subset of the whole ensemble.

The so-called regular octagon, which is the octagon with the highest possible symmetry, where all corners possess an angle of $\pi/4$, was the first being investigated more closely with respect to quantum chaos [8, 72, 73]. Because of the high symmetry one has to desymmetrize the regular octagon in order to compare its statistical properties with RMT. The one-dimensional irreducible representations of the symmetry group corresponds to a hyperbolic triangle with angles $\alpha = \frac{\pi}{2}$, $\beta = \frac{\pi}{3}$ and $\gamma = \frac{\pi}{8}$ having $1/96$ the area of the regular octagon. There are eight possible combinations of Dirichlet and Neumann boundary conditions along the edges $a, b$ and $c$ lying opposite the angles $\alpha, \beta$ and $\gamma$, but only four provide subspectra of the regular octagon. One choice providing a subspectrum of the regular octagon is realized by imposing Dirichlet boundary conditions on all three edges, as considered in [72, 74]. Here we consider two particular combinations of Dirichlet and Neumann boundary conditions, of which one is contained in the spectrum of the regular octagon, called triangular billiard $\mathcal{B}$ in the following, and the other, called triangular billiard $\mathcal{A}$, is not. Both have Neumann boundary conditions on side $c$, but billiard $\mathcal{B}$ has Dirichlet b.c. on side $b$ and Neumann b.c. on side $a$, whereas billiard $\mathcal{A}$ has Dirichlet b.c. on side $a$ and Neumann b.c. on side $b$. The statistical analysis of both billiards is discussed in [11]. The reflection group which generates triangle $\mathcal{B}$ belonging to the regular octagon is arithmetic and thus implies exponentially increasing multiplicities in the length spectrum of periodic orbits. Arithmetical quantum chaotic systems are now known to violate the RMT behaviour even for short-range correlations [9, 10, 11, 12, 13], and the triangular billiard $\mathcal{B}$ shows all features of arithmetic systems. The triangular billiard $\mathcal{A}$ and the other triangular billiards with boundary conditions which do not belong to the regular octagon, show no peculiarities in the quantal spectra although they have nevertheless the same degenerate spectrum of periodic orbits however with different characters attached to the periodic orbits. The following statistics are based on the first 1000 quantal levels for each triangle.

In addition to these two triangles we also investigate an asymmetric hyperbolic octagon which only possesses one involutary symmetry which is always present in hyperbolic octagons. It is not arithmetical and can easily be desymmetrized with respect to the involutary symmetry. The statistical analysis concerns the subspectrum belonging to the positive parity class comprising the first 1127 quantal levels.

## VII.2  Euclidean billiards

In this section examples of two-dimensional Euclidean billiards will be introduced. They are given by the free motion of a point particle inside a compact domain $\Omega$, with elastic reflections at the boundary.

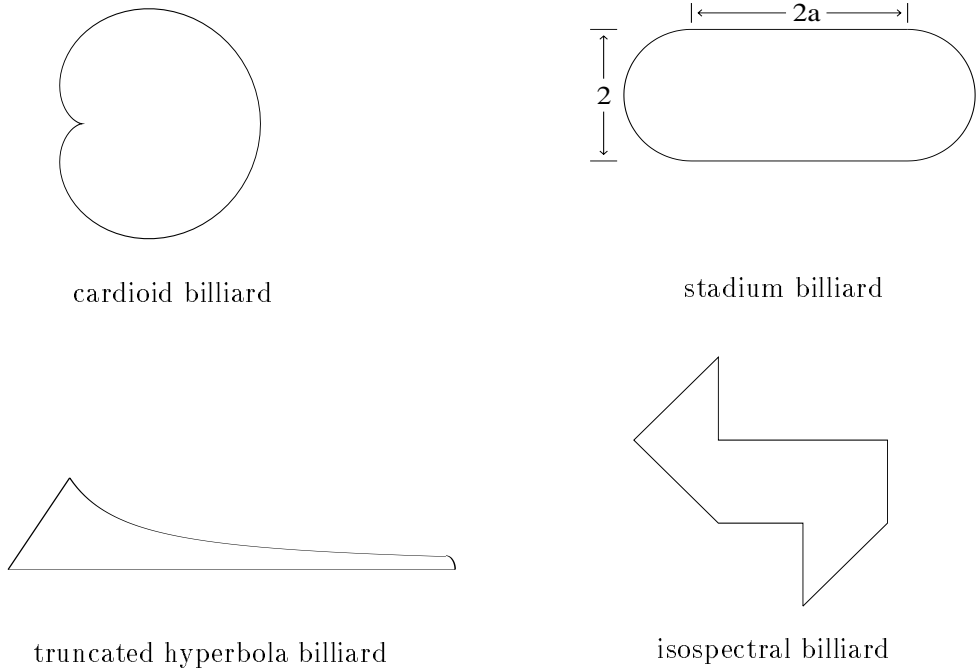

Figure 4: Cardioid billiard, stadium billiard, truncated hyperbola billiard and one of the isospectral billiards.

1. Cardioid billiard

    The boundary curve of the cardioid billiard (see fig. 4) is given in polar coordinates by $\rho(\phi) = 1+\cos\phi$. The cardioid billiard is the limiting case of a family of billiards introduced in [75, 76], whose boundary is defined by a conformal mapping of the unit circle. For the cardioid billiard it is proven that it is ergodic, mixing, a $K$-system and a Bernoulli system [77, 78, 79, 80]. Due to the reflection symmetry of the cardioid billiard, one can divide the quantum mechanical wave functions into odd and even eigenfunctions. In [28] no observable differences between the statistics of the odd and even energy spectra are found (this is also observed in [81]). Thus we use in the following analysis the first 10 000 energy levels for odd symmetry (for a detailed investigation of the spectral statistics for both symmetry classes see [28, 42]). The set of eigenvalues was calculated [82] by means of the conformal mapping technique [76, 83].

    $\overline{N}(E)$, eq. (4), holds with $\mathcal{A} = \frac{3\pi}{4}$, $\mathcal{L} = 6$ and $\mathcal{C} = \frac{3}{16}$ for odd symmetry. There are no prominent slowly varying long range correlations observed [28], such that $N_{\text{long}}(E) = 0$.

2. Stadium billiard

    The stadium billiard [18, 19] is given by two semi-circles of radius 1 joined with two parallel lines of length $2a$ (see fig. 4). In our calculations we choose $a = 1.8$. It is proven that the stadium billiard is ergodic, mixing, a $K$-system, and a Bernoulli system [18, 19].

    We will consider the completely desymmetrized stadium billiard with everywhere Dirichlet boundary conditions, for which we calculated the first 6000 eigenvalues using the boundary element method. The constants in eq. (4) for a quarter stadium with Dirichlet boundary conditions are given by $\mathcal{A} = a+\frac{\pi}{4}$, $\mathcal{L} = 2+\frac{\pi}{2}+2a$ and $\mathcal{C} = \frac{11}{48}$. The additional term $N_{\text{long}}(E)$ is not zero because of the so-called bouncing-ball orbits. These are a family of neutral

periodic orbits bouncing perpendicularly between the straight lines. Their contribution to the spectral staircase is given by [20]

$$N_{\text{bb}}(E) = \frac{a}{\pi} \sum_{n=1}^{\infty} \sqrt{E - E_n^{\text{bb}}} \, \Theta\left(\sqrt{E} - \sqrt{E_n^{\text{bb}}}\right) - \left(\frac{a}{4\pi} E - \frac{1}{2\pi}\sqrt{E}\right) \;,$$

where $E_n^{\text{bb}} = \pi^2 n^2$ are the eigenvalues of a particle in a one-dimensional box of length 1, and $\Theta$ is the Heaviside step function.

There are further contributions from the bouncing-ball orbits due to edge effects [20, 21]. The integrated results of the formulas derived in [20] are

$$\begin{aligned} N_{\text{e1}}(E) &= -\frac{1}{2\pi}\cos(2\sqrt{E}) \\ N_{\text{e2}}(E) &= -\frac{1}{4\pi}\sum_{n=1}^{\infty}\frac{\sin(2n\sqrt{E})}{n} = \begin{cases} 0 & ; \sqrt{E} = \pi \\ \frac{1}{4}\left[\frac{\sqrt{E}}{\pi} \bmod 1\right] - \frac{1}{8} & ; \sqrt{E} \neq \pi \end{cases} \end{aligned}.$$

Both contributions are bounded and thus play no role for the limit distribution for $E \to \infty$. But since we have only a finite number of energy levels, we find a slight improvement if we take them also into account in $N_{\text{long}}(E)$.

3. Truncated hyperbola billiard

   The truncated hyperbola billiard (see fig. 4) is a compact model of the unbounded hyperbola billiard, which is studied in detail in [84]. The boundary is given by the positive $x$-axis, the diagonal $y = x$ and the hyperbola $y = 1/x$, which is truncated at $A = (a, 1/a)$ by a circular arc in such a way that the first derivative is continuous and the arc is perpendicular to the $x$-axis. Using the boundary element method the first 1851 eigenvalues with $E < 10\,000$ are calculated for the truncation parameter $a = 7.1175806$, which corresponds to the value chosen in a recent microwave experiment [85].

   The spectral staircase function displays a prominent long-range modulation, which can be explained by an additional term in Gutzwiller's trace formula [22]. This contribution arises from families of closed, non-periodic orbits which are reflected in the point $A$, where the curvature is discontinuous. This contribution tends to zero for $E \to \infty$, but is necessary for the numerical computation due to the finite number of available energy levels. Thus $N_{\text{long}}(E)$ is given by this additional term. For the exact formula we refer the reader to [22]. There are no proofs concerning the ergodic properties, but numerically the system behaves like a chaotic one.

4. Isospectral billiard

   In 1966 Kac posed the famous question "Can one hear the shape of a drum?" [86]. Since Weyl's law determines the asymptotic behaviour of the spectrum in terms of geometric quantities like the area, circumference and properties of the corners of the membrane, the question arises whether there exist two billiards with identical spectra but different boundaries. Surprisingly there are indeed isospectral billiards as is shown in [87], where two planar isospectral billiards are constructed, each consisting of seven triangles with lengths $(1, 1, \sqrt{2})$. One of them is shown in figure 4. Thus Kac's question is answered in the negative. The first 25 frequencies are determined experimentally in [88]. Using a mode-matching method the first 598 quantal levels are computed in [89]. It is demonstrated that the spectral statistics lies between the GOE and the Poisson behaviour with respect to the short-range correlations. However, when the quantal spectrum belonging to the triangle is subtracted, an agreement with GOE is observed. We do not pursue this approach because the spectrum of the triangle is not a subspectrum of the isospectral billiard since it cannot be obtained by desymmetrization.

Using the boundary element method the quantal levels $E_n$ in the low-energy range $E_n \in [0, 5000]$ and a high-energy range $E_n \in [37000, 45000]$, have been computed. The first energy range comprises of 1335 quantal levels, the second one of 2210 levels.

5. Triangular billiard

   Since the isospectral billiard is built up from triangles with lengths $(1, 1, \sqrt{2})$, we want to study the eigenvalue statistics of one of these integrable triangles separately. The eigenfunctions are given by

   $$\Psi_{m,n}(x,y) = \frac{1}{\sqrt{2}} \left[ \sin(\pi m x) \sin(\pi n y) - \sin(\pi n x) \sin(\pi m y) \right]$$

   where $n, m \in \{1, 2, \ldots\}$ with $n < m$. Thus the eigenvalues are

   $$E_{n,m} = \pi^2 \left( n^2 + m^2 \right) \quad ; \quad n, m \in \{1, 2, \ldots\}, \text{ with } n < m \ .$$

   The corresponding constants of the generalized Weyl's law in eq. (4) are given by $\mathcal{A} = \frac{1}{2}$, $\mathcal{L} = 2 + \sqrt{2}$ and $\mathcal{C} = \frac{3}{8}$. Altogether the first $10^6$ eigenvalues are used in the following calculations.

   Note that for this system a non-Gaussian behaviour for the mode-fluctuation distribution $P(W)$ holds, see section VI.

6. Circular billiard (even symmetry)

   In case of the integrable unit circular billiard the (non-normalized) eigenfunctions are given in polar coordinates by

   $$\Psi^{\pm}_{m,n}(r, \varphi) = J_m(k_{m,n} r) e^{\pm i m \varphi} \qquad \text{with} \ \ m = 0, 1, 2, \ldots, \ m = 1, 2, 3, \ldots \qquad .$$

   The boundary condition $\Psi^{\pm}_{m,n}(1, \varphi) = 0$ leads to the eigenvalues $E_{m,n} = k^2_{m,n}$ as the squares of the positive zeros $k_{m,n}$ of the Bessel functions, i.e., $J_m(k_{m,n}) = 0$. For $m \neq 0$ the solutions are twofold degenerate; we will consider the case of the desymmetrized circular billiard with even symmetry with respect to the reflection at the diameter. The set of eigenvalues is given by $\{E_{m,n}\}$, and each eigenvalue occurs with multiplicity one. The corresponding constants in eq. (4) are given by $\mathcal{A} = \frac{\pi}{2}$, $\mathcal{L} = \pi - 2$ and $\mathcal{C} = -\frac{1}{24}$. Altogether the first $10^5$ eigenvalues for even symmetry are used in the following calculations.

7. Torus

   As a further integrable system we study a system which is defined on the two-dimensional torus (see [27, 50]). The eigenvalues of the "displaced" Laplace operator

   $$\hat{H} = (i\vec{\nabla} + \vec{\alpha})^2 = -(\partial_x^2 + \partial_y^2) + \alpha_1^2 + \alpha_2^2 + 2i\alpha_1 \partial_x + 2i\alpha_2 \partial_y \ ,$$

   with $(x, y) \in [0, 2\pi]^2$ together with periodic boundary conditions read

   $$E_{n_1, n_2} = (n_1 - \alpha_1)^2 + (n_2 - \alpha_2)^2 \quad ; \quad n_1, n_2 \in \mathbf{Z} \ , \ \alpha_1, \alpha_2 \in [0, 1[ \ .$$

The eigenfunctions are given by

$$\Psi(x,y) = e^{in_1 x} e^{in_2 y} \quad .$$

In the following analysis we use the first $10^6$ eigenvalues for the case $\alpha = (0.3437, 0.4304)$, which produces a bimodal distribution of $P(W)$ [27, 50, 62].

For this class of systems it is proven in [50] that the corresponding limit distribution is non-Gaussian. The case with $\alpha_1 = \alpha_2 = 0$ was treated before in [59].

8. Rectangular billiard

The last integrable system used in our comparison between chaotic and non-chaotic systems is a rectangular billiard with ratio $a/b = 5/7$, and with Dirichlet boundary conditions. The (non-normalized) eigenfunctions are given by

$$\Psi_{m,n}(x,y) = \sin\left(\frac{\pi m x}{a}\right) \sin\left(\frac{\pi n y}{b}\right) \quad , \quad m,n \in \mathbf{N} \quad . \tag{71}$$

The corresponding eigenvalues are $E_{m,n} = \pi^2 \left(\frac{m^2}{a^2} + \frac{n^2}{b^2}\right)$. This system is chosen because of a recent experiment using such a microwave cavity [90]. In the experiment the first 657 eigenvalues were measured and it is found that the mode-fluctuation distribution $P(W)$ is very close to a Gaussian though the system is non-chaotic. Here we base our analysis on the first $10^6$ eigenvalues, demonstrating clear deviations from the Gaussian, see below. Notice, that also for this class of systems a non-Gaussian behaviour of the mode-fluctuation distribution $P(W)$ is proven, see section VI.

## VII.3 Numerical comparison of chaotic and non-chaotic systems

Let us now turn to the numerical comparison of the chaotic and non-chaotic systems which we have introduced above. In the numerical computations we have calculated a superposition of the statistics for different values of $T$, such that all available eigenvalues are taken into account. At first figure 5 shows the $E(k,L)$ functions in the case of the cardioid billiard using the first 10 000 quantal levels of the odd symmetry class in comparison with the GOE behaviour. The range $L = 0\ldots 10$ is displayed in figure 5a, and $L = 50\ldots 60$ in figure 5b. One observes a good agreement with the GOE behaviour up to $L \simeq 3$. Thereafter the width of the $E(k,L)$ functions saturates, whereas the width of the GOE curves increases logarithmically as discussed in section III.2. The increasing width corresponds to a decreasing maximal height of the (nearly perfect) Gaussians which is exemplified in figure 5b where the GOE curves are lying below the curves belonging to the cardioid billiard. This means that the spectrum of the cardioid billiard is at large correlation lengths more rigid than a GOE spectrum. The saturation effect is more clearly visible in figure 6 where the range $L = 0\ldots 60$ is shown. The variations in the widths and in turn in the amplitudes are determined by the number variance $\Sigma^2(L)$, see eq.(34), which fluctuates around the saturation plateau.

In order to check the agreement of the $E(k,L)$ functions with a Gaussian, we map the $E(k,L)$ functions onto each other. To that end we apply the substitution $\ell = \frac{L-c}{\sigma}$, where $c$ and $\sigma$ denote the mean value and the standard deviation of the $E(k,L)$ functions, i.e., $E^{\mathrm{n}}(k,\ell)d\ell = E(k,L)dL = E(k,\ell\sigma + c)\sigma d\ell$. The transformed $E(k,L)$ functions are displayed in figure 7 for the cardioid billiard in a), the stadion billiard in b), the truncated hyperbola billiard in c), the hyperbolic octagon in d), the hyperbolic triangle $\mathcal{A}$ in e) and the arithmetic hyperbolic triangle $\mathcal{B}$ in f). In these figures the $E(k, L)$ functions from $k = 75$ up to $k = 125$ are mapped onto each other. In all cases a Gaussian with variance $\sigma^2 = 1$ lies well in the cloud of the $E(k, L)$ functions, indicating the good agreement with a Gaussian. Nevertheless, there are system dependent differences as a comparison between the cardioid billiard with the hyperbolic triangle $\mathcal{B}$ shows.

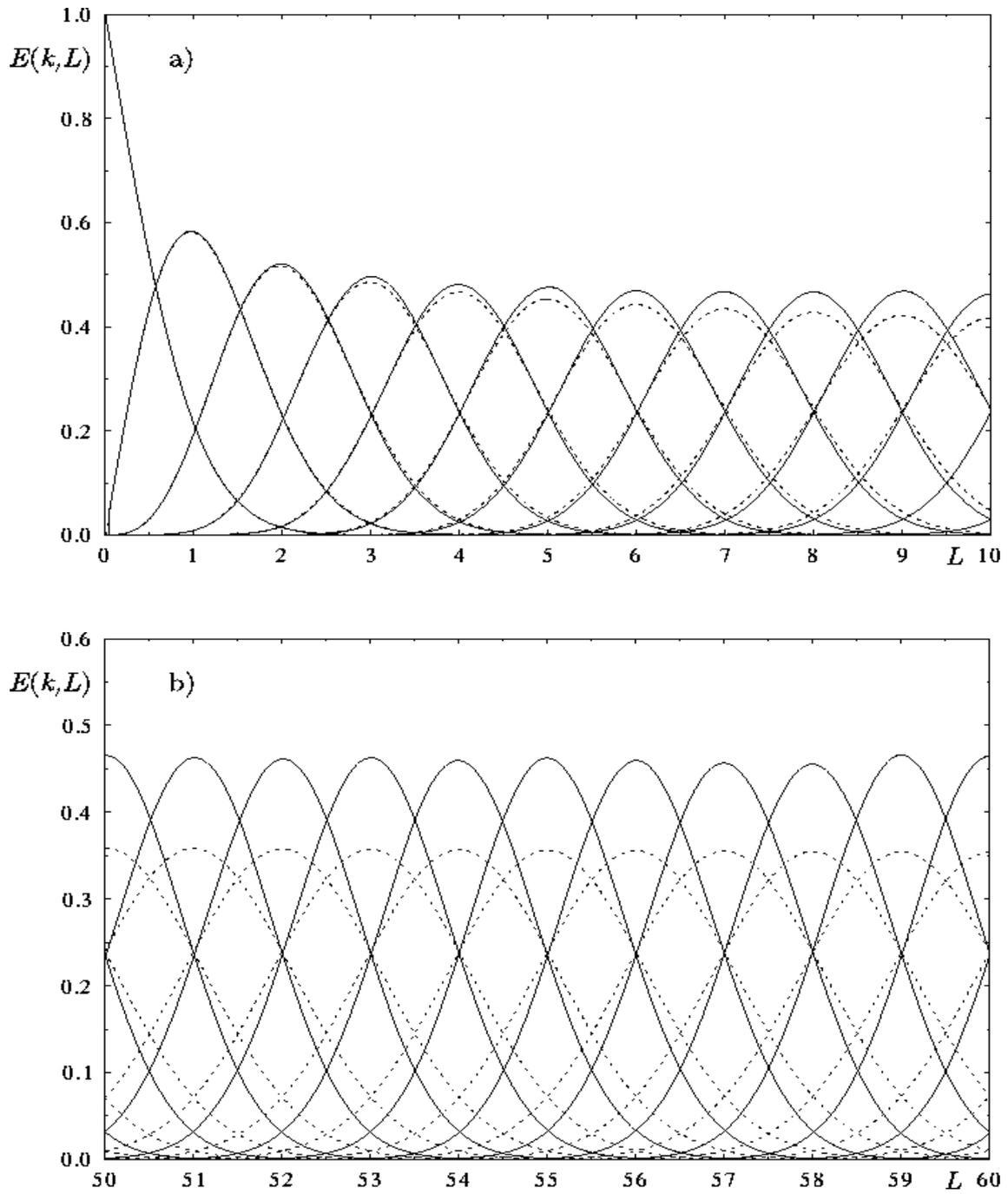

Figure 5: The $E(k,L)$ functions are shown for the cardioid billiard using the first 10 000 odd quantal levels (full curves) in comparison with the GOE behaviour (dotted curves). The two ranges $L = 0 \ldots 10$ and $L = 50 \ldots 60$ are displayed.

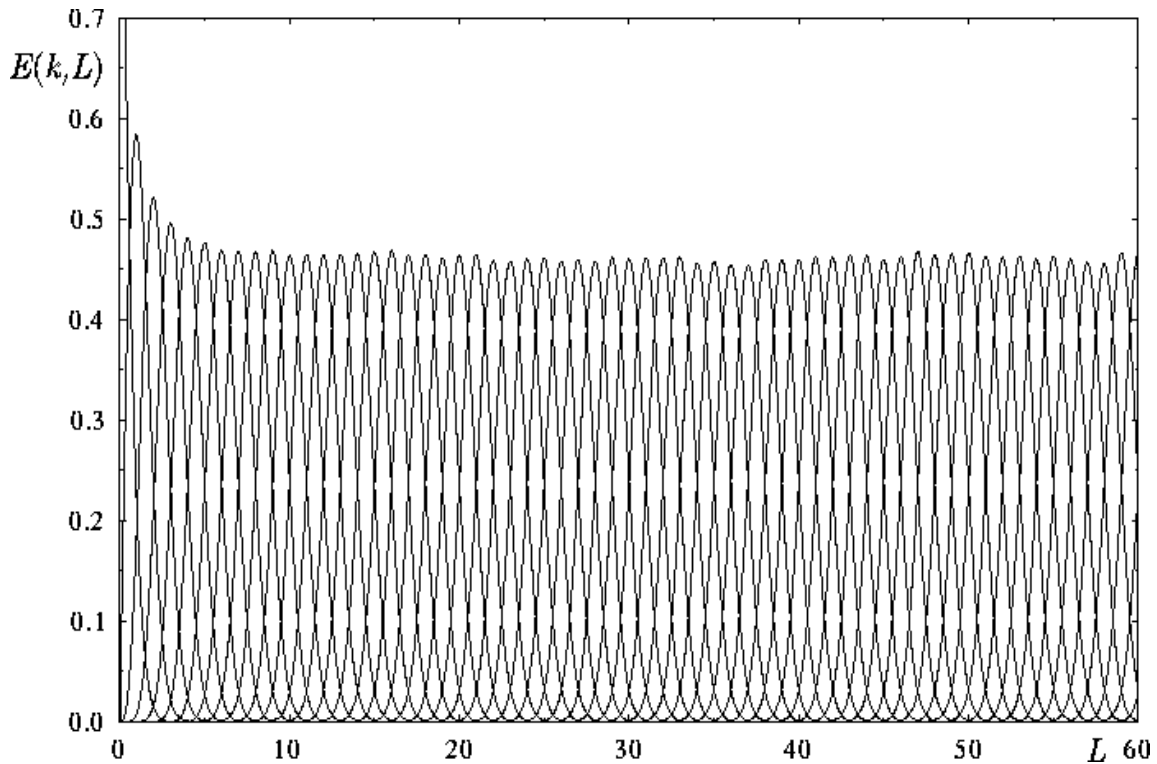

Figure 6: The $E(k, L)$ functions are shown for the cardioid billiard using the first 10 000 odd quantal levels in the range $L = 0 \ldots 60$.

In figure 8 the same mapping is applied to five non-chaotic systems again for $E(k, L)$ from $k = 75$ up to $k = 125$ in the case of the isospectral billiard, and from $k = 5000$ up to $k = 5050$ for the other systems. The high $k$ values are necessary because the latter four systems reach their non-universality regime much later than $k = 125$. Being in the non-universality regime is, however, required for a comparison with the mode-fluctuation distribution $P(W)$. The results for the isospectral billiard are displayed in figure 8a and 8b for the low and a high energy range, respectively, as discussed in the previous section. It is seen that the distributions are not stationary since they change significantly from the low-energy to the high-energy range, which is in contrast to the case of chaotic systems where the Gaussian is approached very fast. Figure 8c displays the $E(k, L)$ functions for the Euclidean triangle billiard, figure 8d for the circular billiard, figure 8e for the torus and figure 8f for the rectangular billiard.

The $E(k, L)$ functions for non-chaotic systems are found to be neither Gaussian nor Poissonian. This is important because if they were Poissonian they would also tend to a Gaussian in the high $k$ limit, and thus lead to a Gaussian behaviour of $W(x)$ as we have seen in section IV. Only $E(k, L)$ functions showing a non-Gaussian structure for $L \to \infty$ can generate a non-Gaussian $P(W)$.

A possible qualitative measure for the deviations from the normalized Gaussian behaviour

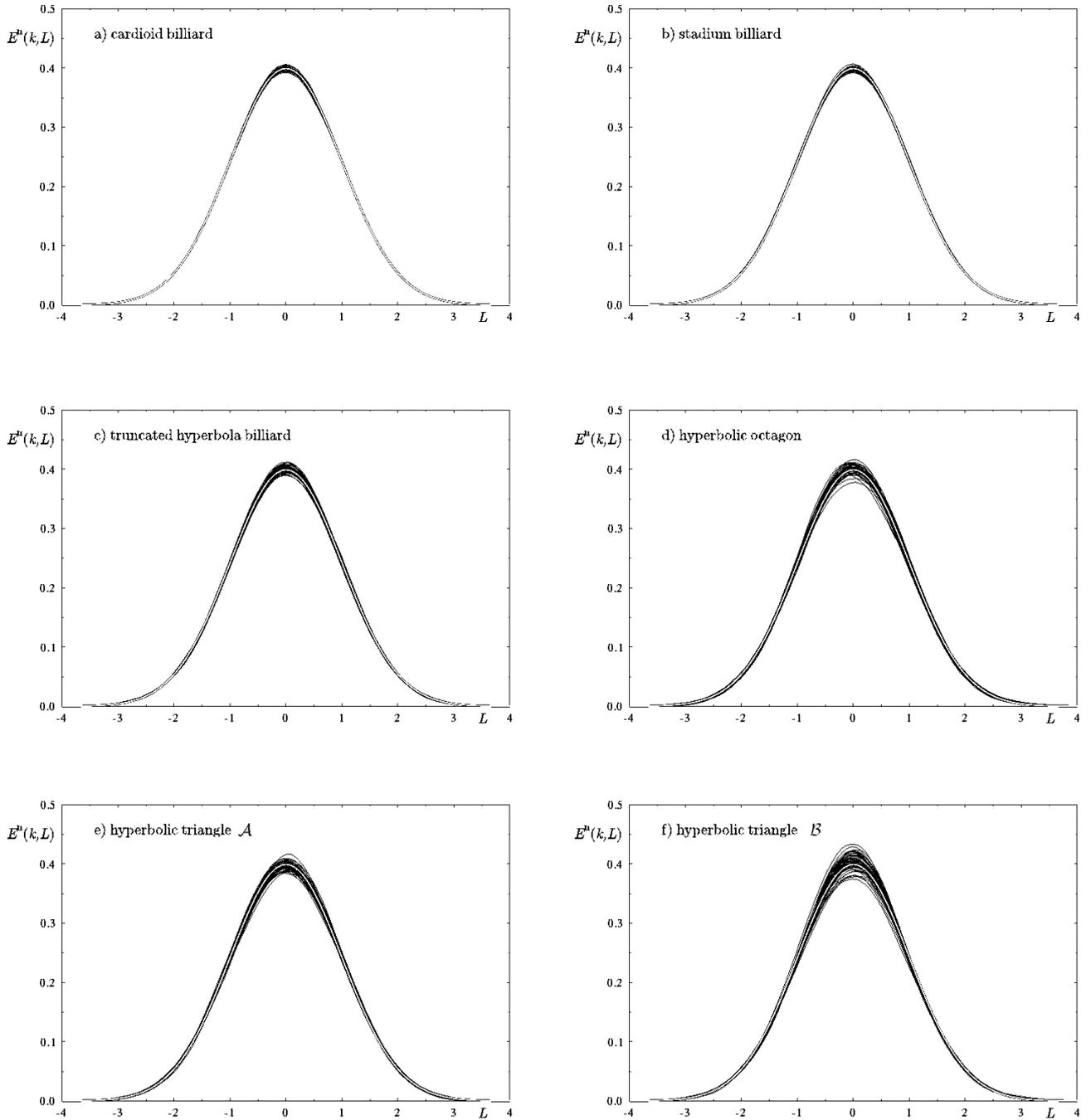

Figure 7: For six chaotic systems the $E(k, L)$ functions are normalized as discussed in the text for $k = 75$ up to 125 such that they are mapped onto each other. The curves are in good agreement with the normalized Gaussian shown as white curve.

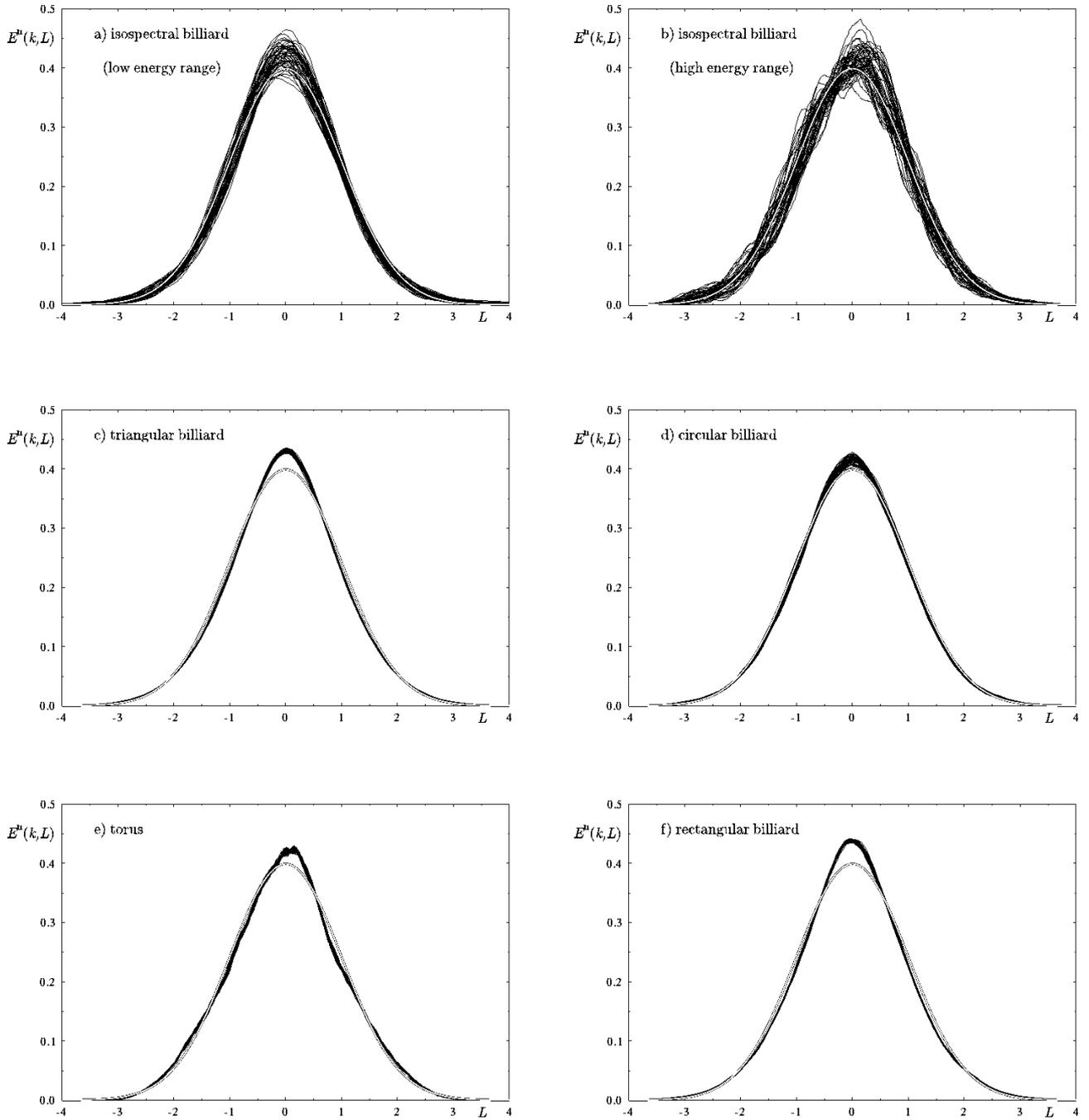

Figure 8: For five non-chaotic systems the $E(k, L)$ functions are normalized as discussed in the text such that they are mapped onto each other. The $k$-values are chosen out of the saturation range. Two distinct energy ranges are shown for the isospectral billiard with $k = 75..125$. For the other systems we have chosen $k = 5000..5050$. The curves show clear deviations from the normalized Gaussian shown as white curve.

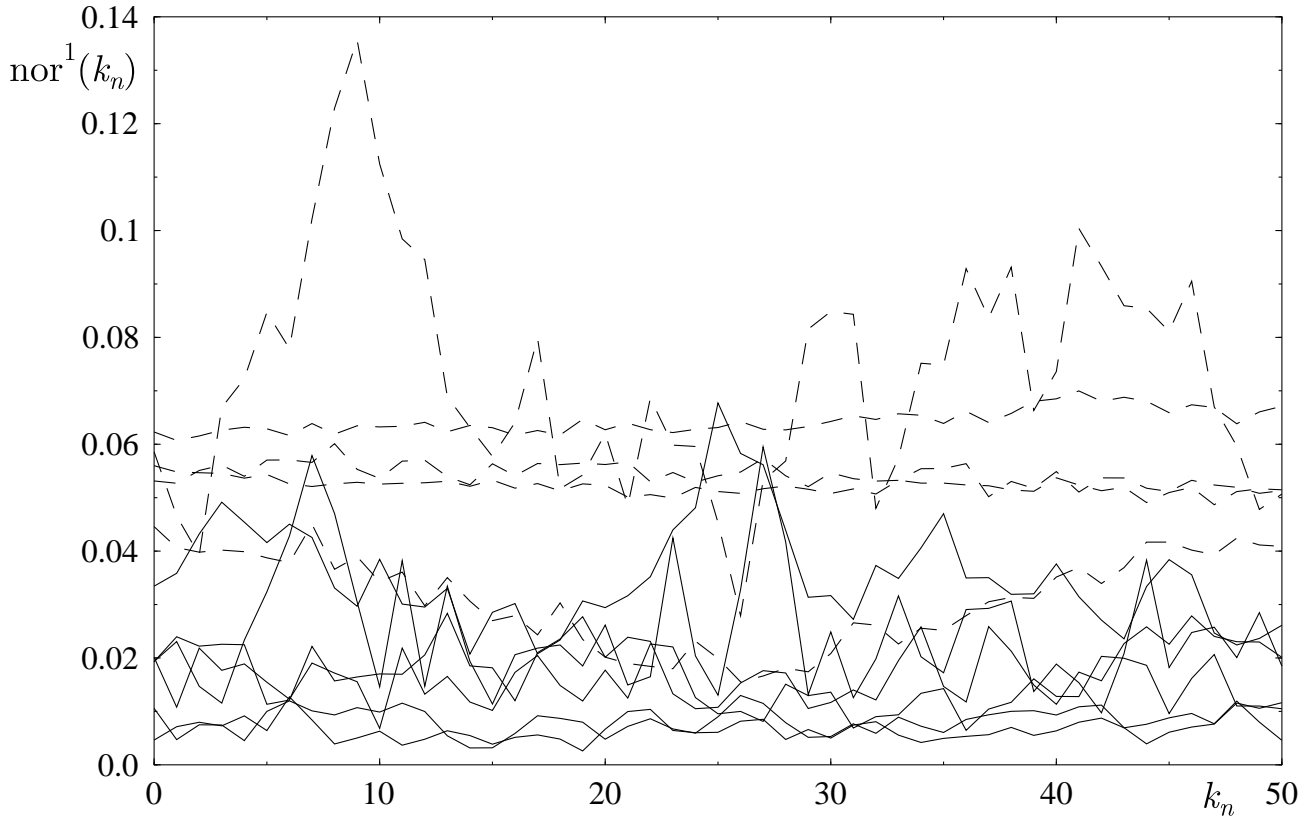

Figure 9: The $q = 1$ norms are shown for the chaotic (solid curves) and for the non-chaotic systems (dashed curves) using the ordinate $k_n = k - k_d$ as explained in the text.

is provided by the following $L^q$-norm

$$\text{nor}^q(k) := \left( \int_{-\infty}^{\infty} d\ell \left| E^n(k,\ell) - \frac{1}{\sqrt{2\pi}} e^{-\frac{\ell^2}{2}} \right|^q \right)^{1/q} . \qquad (72)$$

We have tested different $L^q$-norms and found that the norm for $q = 1$ shows the differences between chaotic and non-chaotic systems slightly clearer than higher $L^q$-norms. In figure 9 the results for our systems are shown. To display all the $\text{nor}^q(k)$'s in a single figure, the $k$-values have to be shifted according to $k_n = k - k_d$, i.e., for the chaotic systems as well as for the isospectral billiard we subtract $k_d = 75$ and for the integrable systems $k_d = 5000$ from the $k$-values. Using the ordinate $k_n = k - k_d$, the norms of the chaotic systems are plotted in figure 9 as solid curves and of the non-chaotic systems as dashed curves. One observes a clear tendency towards larger values for the non-chaotic systems. However, there is the exception of the circular billiard whose norm is at some values of $k_n$ of the same order as for chaotic systems. On the other hand, at $k_n \simeq 25$, the $L^1$-norm of the hyperbolic octagon and the triangle $\mathcal{B}$ display values ranging into the order of non-chaotic ones. But for the other non-chaotic systems the $L^q$-norm yields a possible measure to discriminate between a Gaussian and a non-Gaussian behaviour.

Another measure is provided by the moments and cumulants of $\eta_L(x)$ which can be expressed in terms of the $E(k, L)$ functions having the advantage that they catch different aspects of the distribution separately. To emphasize this point, figures 10 and 11 show the seven lowest cumulants of $\eta_L(x)$ $C_k(L)$ for the same systems as in figures 7 and 8. Here the cumulants have been smoothed in order to stress their mean behaviour. Because of the normalization, the second cumulant $C_2$ is identical to one. In the case of the chaotic systems all considered cumulants fluctuate around zero which is in contrast to the non-chaotic case where some cumulants $C_k$ have large non-vanishing values revealing systematic deviations from the Gaussian behaviour. E. g., one observes for the triangular, the circular and the rectangular billiard large non-zero values for $C_4$ and $C_6$. In other cases more pronounced fluctuations occur than for chaotic systems.

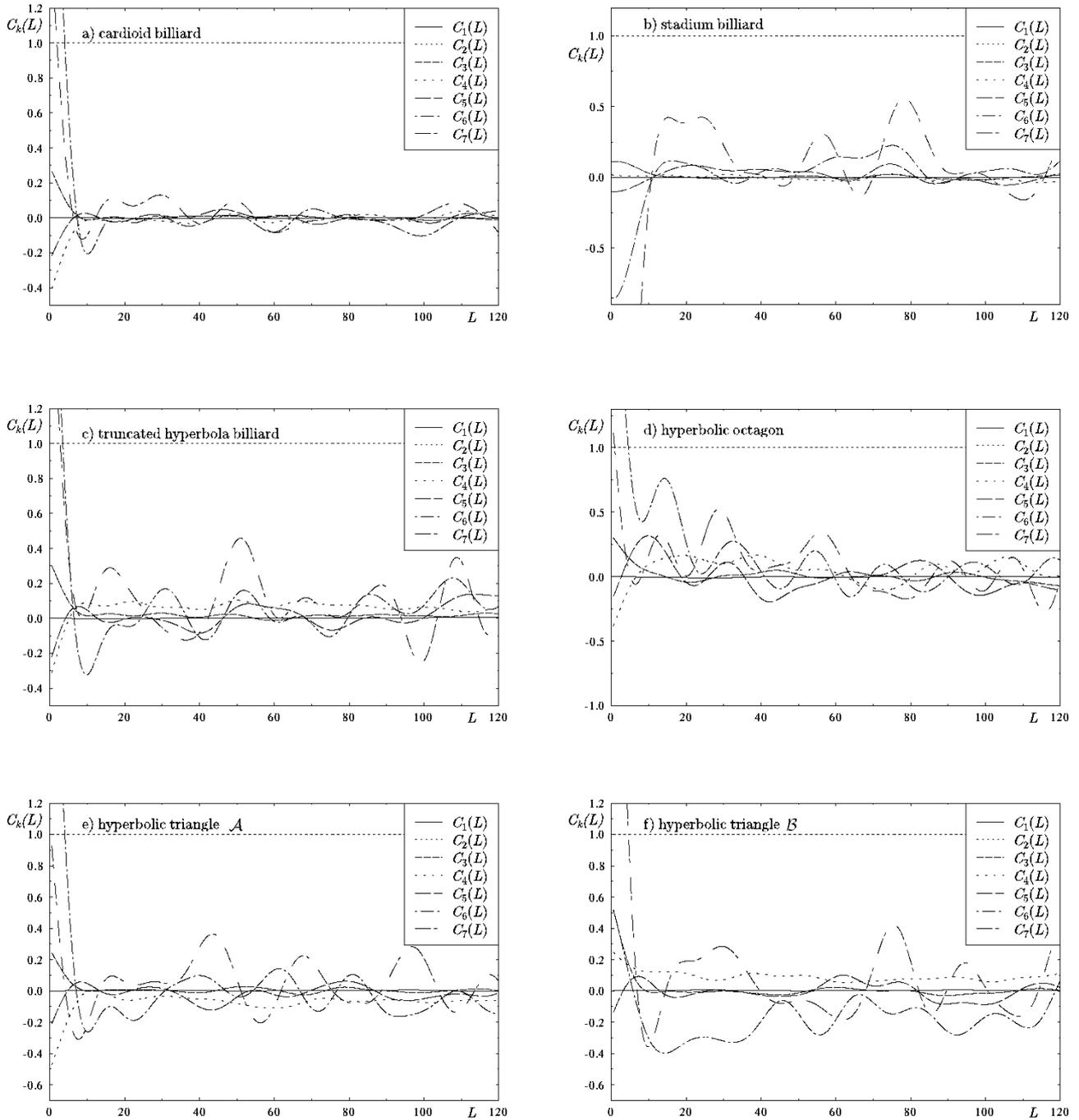

Figure 10: The first seven smoothed cumulants $C_k(L)$ are displayed for six chaotic billiards.

The cumulants of $\eta_L(x)$ are computed from the moments which in turn are computed from

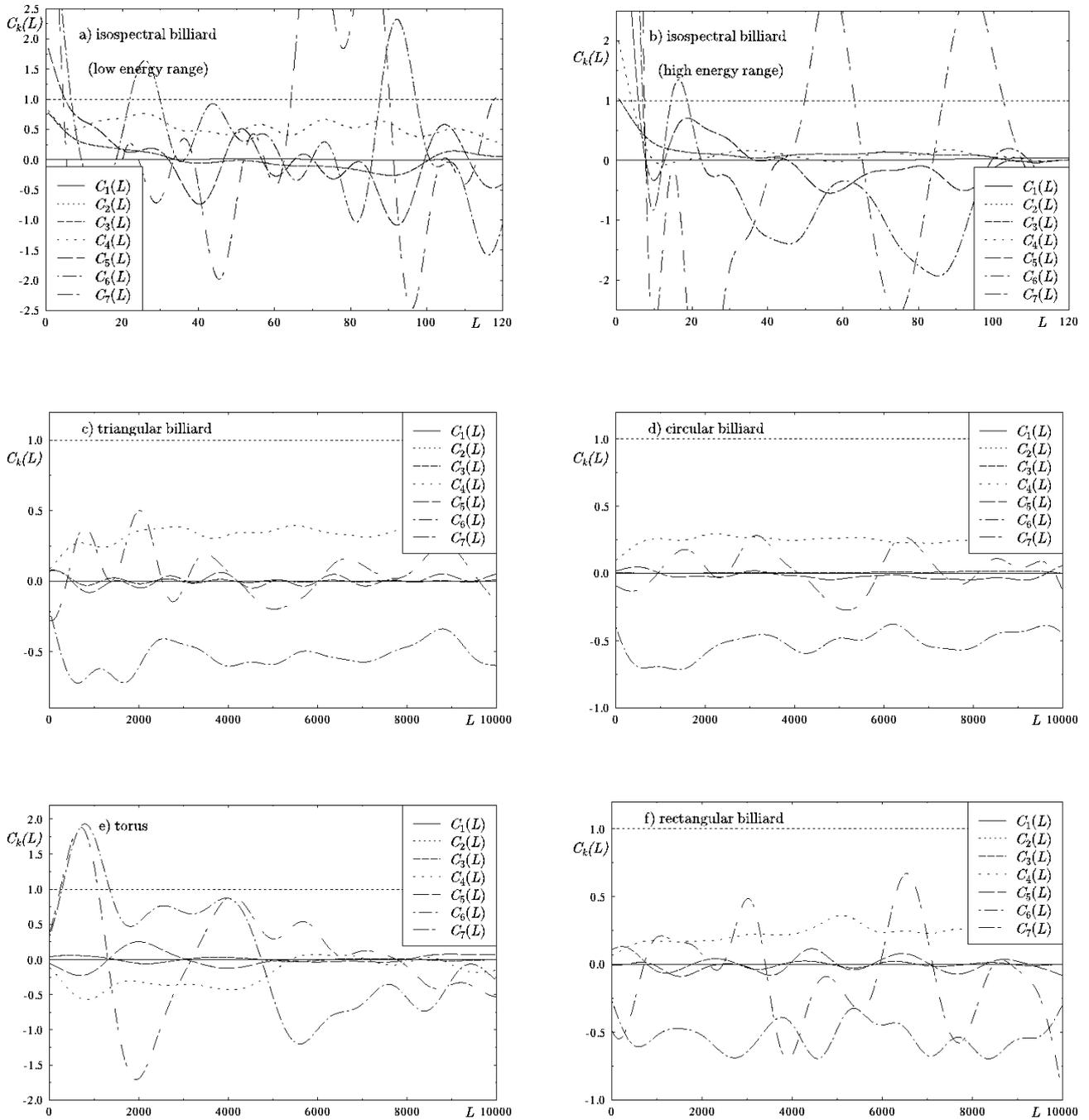

Figure 11: The first seven smoothed cumulants $C_k(L)$ are displayed for five non-chaotic billiards. Two distinct energy ranges are shown for the isospectral billiard.

the $E(k,L)$ functions using

$$m_j(L) = \sum_{k=0}^{\infty} k^j \, E(k,L) \quad . \tag{73}$$

The cumulants are then obtained by means of (48)

$$c_k(L) = \left. \frac{d^k}{d\mu^k} \ln M(\mu) \right|_{\mu=0} = \left. \frac{d^k}{d\mu^k} \ln \left( \sum_{j=0}^{\infty} m_j(L) \frac{\mu^j}{j!} \right) \right|_{\mu=0} , \tag{74}$$

where the series expansion of the logarithm yields the dependence of $c_k(L)$ on $m_j(L)$ with $j \leq k$ [44]. From this one obtains (see eq.(47)) the cumulants of $\eta_L(x)$, i.e., $C_1 = (c_1 - L)/\sqrt{c_2}$ and $C_k = c_k/c_2^{k/2}$ for $k > 1$.

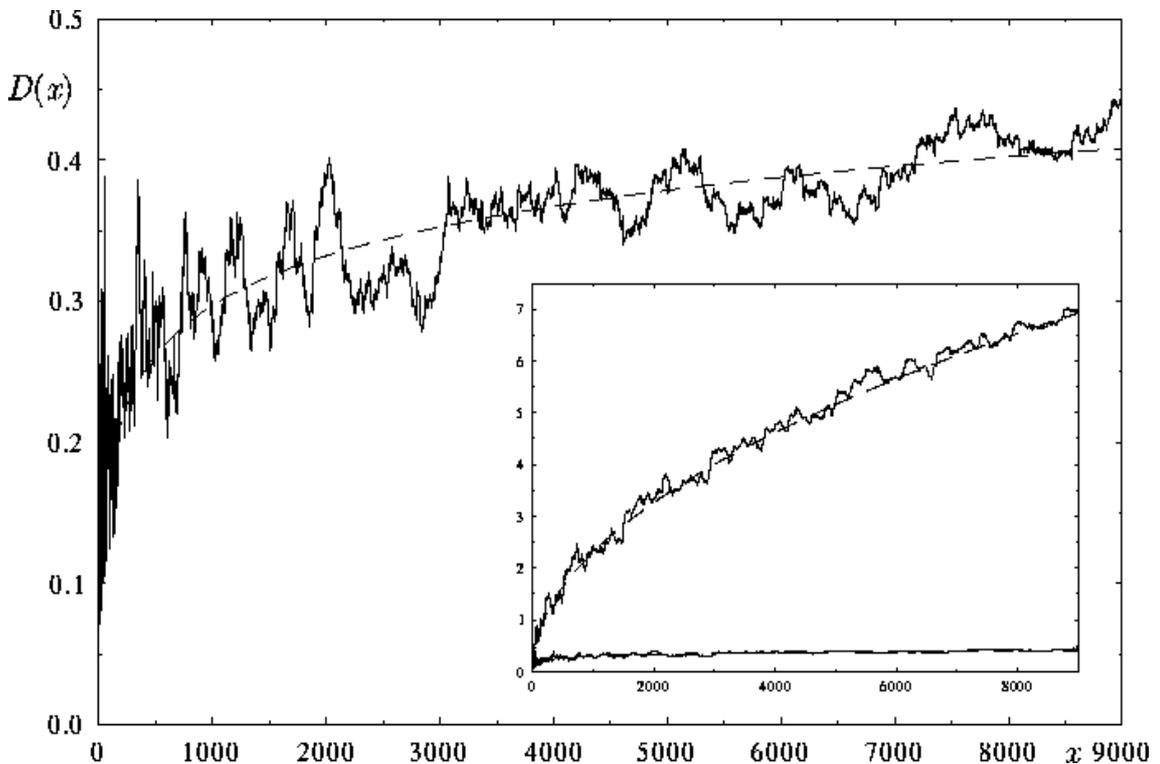

Figure 12: $D(x)$ is shown for the cardioid billiard using $c = 1.1$. The inset shows the same curve in comparison with that for the circular billiard where the increase is proportional to $\sqrt{x}$.

Before we finally turn to the $P(W)$ distribution, we present in figure 12 $D(x)$ for the cardioid billiard, using the first 10 000 levels and $c = 1.1$. In the inset additionally $D(x)$ of the integrable circular billiard is shown. The first possesses a slowly logarithmically increasing $D(x)$, whereas the latter displays an increase proportional to $\sqrt{x}$. We have checked that the dependence on $c$ is very weak. This provides us with a convenient method for computing the distribution $P(W)$ in contrast to earlier methods. These earlier methods required, firstly, the determination of the spectral rigidity and its saturation value $\Delta_\infty(E)$ in dependence on the energy $E$, and secondly, a fit to $\Delta_\infty(E)$ in order to obtain its energy dependence [11, 15, 28]. This fit function yields an approximation to the variance $D(x)$ of the mode fluctuations. Using the definition (52) for $D(x)$, one obtains $W(x)$ solely depending on the parameter $c$. From a numerical point of view, this parameter should be close to one since the larger the value of $c$ the larger is the part of the spectrum being excluded from the analysis of $P(W)$. However, a value too close to one implies an averaging over a statistically too small interval. The optimal value of $c$ is thus a compromise depending on the number of available quantal levels. We have tested values in the range $c = 1.1 \ldots 1.5$ and observe only slight differences in the mode-fluctuation distribution $P(W)$. In figures 13 and 14 the distribution $P(W)$ is shown for the same systems as in figures 7 and 8. Here $c = 1.1$ has been used, with the exception of the hyperbolic octagon where $c = 1.5$ has been used. This is due to the observation that this system shows a $c$-dependence at $c = 1.1$, which can occur if not enough quantal levels are available. The difference in the mode-fluctuation distribution $P(W)$ between the chaotic and non-chaotic systems is obvious. The chaotic systems display a very good agreement with the Gaussian behaviour, whereas significant deviations are observed in the case of the non-chaotic systems.

|  | skewness | kurtosis |
|---|---|---|
| cardioid billiard | -0.018 | -0.14 |
| stadium billiard | 0.006 | -0.15 |
| truncated hyperbola billiard | -0.036 | -0.12 |
| hyperbolic octagon | 0.015 | -0.18 |
| hyperbolic triangle $\mathcal{A}$ | 0.026 | -0.20 |
| hyperbolic triangle $\mathcal{B}$ | 0.010 | -0.06 |
| isospectral billiard (low energy) | -0.172 | 0.01 |
| isospectral billiard (high energy) | -0.366 | -0.03 |
| triangular billiard | -0.318 | -0.04 |
| circular billiard | 0.233 | -0.12 |
| torus | -0.062 | -0.79 |
| rectangular billiard | -0.218 | -0.14 |

Table 1: The skewness and the kurtosis of the distribution $P(W)$ are listed.

The mean of the distribution $P(W)$ is of the order 0.003 and the variances are accurate to $1 \pm 0.01$ in most cases. In table 1 the skewness and the kurtosis are listed for the 12 mode-fluctuation distributions $P(W)$. Relatively small values are observed for the chaotic systems. The modulus of skewness of the circular, the triangular and the rectangular billiard as well as of the isospectral billiard in the high energy range is larger than 0.2. The large value of the skewness in the case of the integrable circular billiard is also nicely reflected in figure 3 for $N_{\text{fluc}}(x)$, where the fluctuations are larger on the positive side than on the negative one. This leads to a suppression of the tail on the negative side in $P(W)$ in comparison with the Gaussian. For the torus, where the skewness is of the order observed in chaotic systems, the modulus of the kurtosis is with 0.8 very large. Thus these two measures also display the deviations from a Gaussian distribution in the case of non-chaotic systems.

A widely used test for the validity of a statistical hypothesis for discrete (cumulative) distributions is the Kolmogorov-Smirnov test. Since the mode-fluctuation distribution $P(W)$ is a continuous statistic, we apply this test to the distribution of the $\delta_n$, eq.(6), analogously normalized as $W(x)$ but using for the computation of $D(x)$ only the values at the quantal levels. This can be viewed as a discrete version of the distribution $P(W)$ since the fluctuations are only evaluated at energies corresponding to quantal levels. In table 2 the significance niveaus for a Gaussian distribution are shown where for the chaotic and for the isospectral systems all available quantal levels are used. For the remaining integrable systems only the first 10 000 levels are used to employ for the comparison roughly the same order of quantal levels. It is clearly seen that the significance niveaus are much larger in the case of chaotic systems. Using for the integrable systems the first $10^6$ levels leads to significance niveaus of 0%. However, using only the first 1000 levels yields significance niveaus larger than 10% for the circular and for the rectangular billiard. This shows that the deviations from the Gaussian are extremely significant for the integrable systems in figure 14 where $10^6$ level have been used. This is in contrast to the case of chaotic systems where at most $10^4$ levels have been used for the $P(W)$ statistic in figure 13.

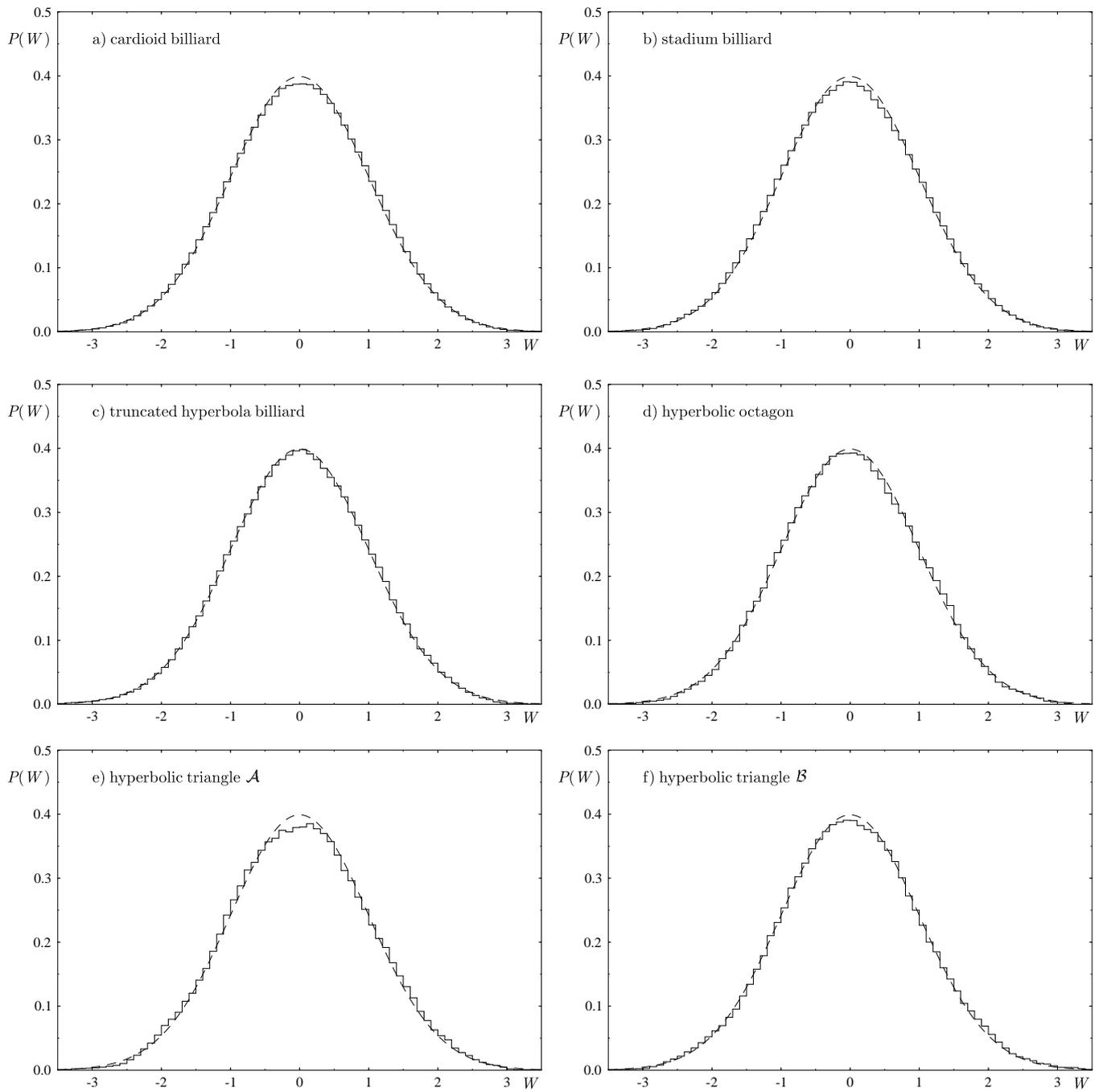

Figure 13: The mode-fluctuation distribution $P(W)$ is displayed for six chaotic billiards.

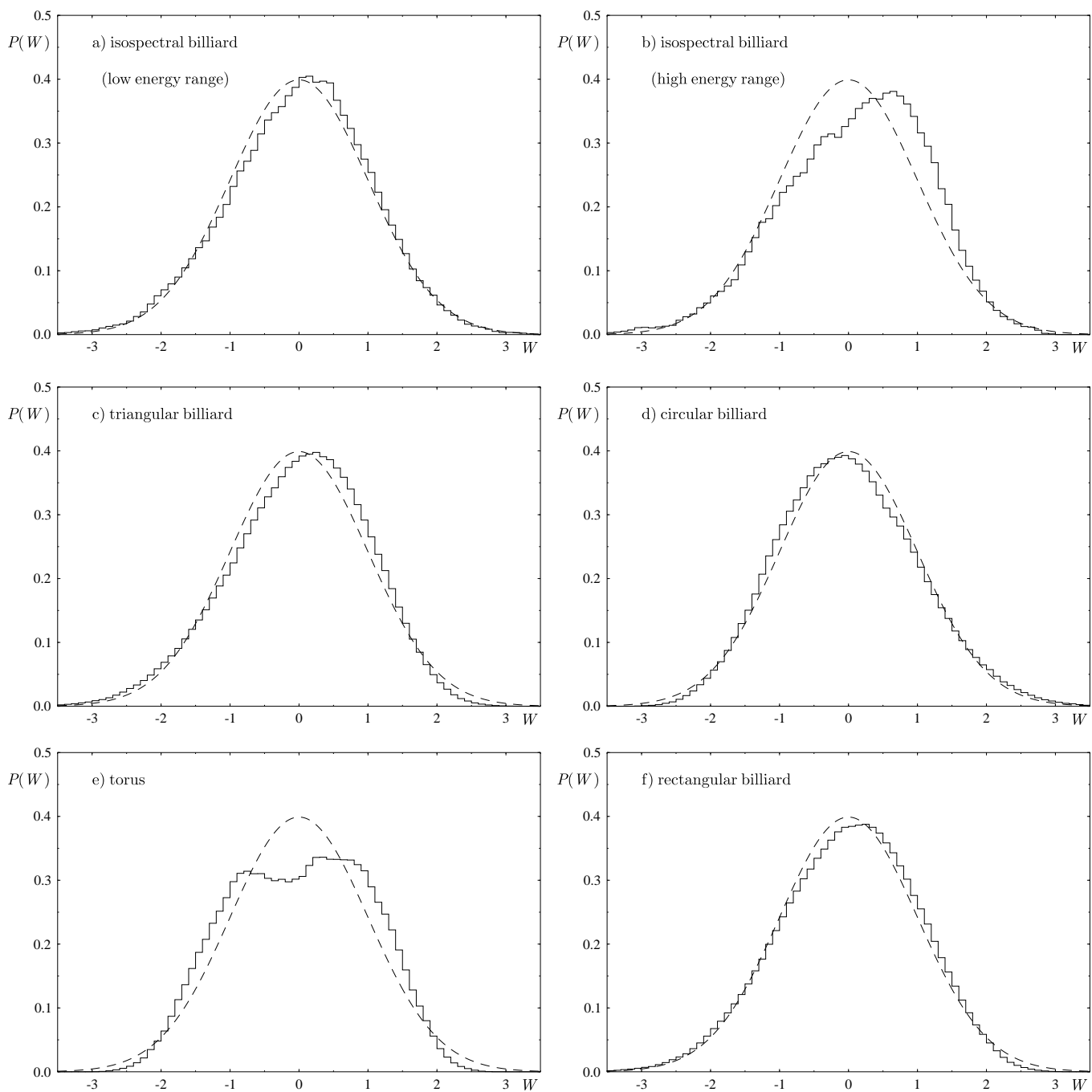

Figure 14: The mode-fluctuation distribution $P(W)$ is displayed for five non-chaotic billiards. Two distinct energy ranges are shown for the isospectral billiard.

|  | significance niveau |
|---|---|
| cardioid billiard | 48% |
| stadium billiard | 77% |
| truncated hyperbola billiard | 99% |
| hyperbolic octagon | 91% |
| hyperbolic triangle $\mathcal{A}$ | 63% |
| hyperbolic triangle $\mathcal{B}$ | 77% |
| isospectral billiard (low energy) | 12% |
| isospectral billiard (high energy) | $10^{-9}$% |
| triangular billiard | $10^{-33}$% |
| circular billiard | 0.02% |
| torus | $10^{-14}$% |
| rectangular billiard | $10^{-5}$% |

Table 2: The significance niveaus of the distribution of the $\delta_n$ with respect to a Gaussian are given.

## VIII  Summary

In this paper we carry out a detailed comparison of chaotic and non-chaotic systems with respect to the statistical properties of their quantal level spectra. Emphasis is put on the $E(k, L)$ functions and the mode-fluctuation distribution $P(W)$.

The RMT properties of the $E(k, L)$ functions are discussed, as well as their behaviour for chaotic and non-chaotic systems. A crucial property is the Gaussian behaviour of the $E(k, L)$ functions in the case of chaotic systems, which in turn leads to a Gaussian distribution of $P(W)$. The $L^1$-norm of the deviation of the $E(k, L)$ functions from the Gaussian shows a clear tendency towards smaller values in the case of the six chaotic billiards, in contrast to the five non-chaotic billiards considered by us. In addition, the cumulants $C_k(L)$ of $\eta_L(x)$ reveal the same property since they fluctuate with small amplitudes around zero for the chaotic systems, whereas at least some cumulants display significant non-zero values or large fluctuations in the case of the non-chaotic systems.

The distribution of $\eta_L(x)$ is studied, which is known to be Gaussian in RMT and also for a Poissonian random process for $L \to \infty$. Furthermore, a Gaussian behaviour of the $E(k, L)$ functions implies a Gaussian behaviour of the distribution of $\eta_L(x)$.

Finally the mode-fluctuation distribution $P(W)$ is studied. There are rigorous proofs that $P(W)$ is non-Gaussian for several integrable systems, leading to the suggestion that they provide a quantity for the distinction between chaotic and non-chaotic systems, because numerical computations show a Gaussian behaviour for chaotic systems. To confirm the conjecture that strongly chaotic systems with only unstable and isolated periodic orbits display a Gaussian $P(W)$ in contrast to non-chaotic ones, we carry out a statistical analysis based on six chaotic and five non-chaotic billiard systems. The distribution of $W(x) := N_{\text{fluc}}(x)/\sqrt{D(x)}$ is numerically shown to be in very good agreement with a Gaussian in the case of the six chaotic systems, whereas for the five non-chaotic systems clear deviations are found.

A Gaussian limit distribution of $P(W)$ for strongly chaotic systems is also satisfactory from the conceptual point of view. The often studied nearest-neighbour level-spacing distribution shows for integrable systems a Poissonian behaviour, in general, leading to the fact that integrable or regular systems possess the most uncorrelated spectra. If one wants to have the degree of classical regularity correspond to the properties of a quantum mechanical statistic, a Poissonian behaviour for regular systems can be seen as surprising. In the case of the normalized mode-fluctuation distribution $P(W)$, despite their rigid spectra, strongly chaotic systems display over long-range correlations the most uncorrelated spectra, which is expressed by the fact that a Gaussian distribution has the largest possible spectral entropy.

Thus our results suggest that a clear fingerprint of quantum chaos is given by a Gaussian behaviour of the mode-fluctuation distribution $P(W)$.

**Acknowledgments**

We would like to thank Jens Bolte for many useful discussions and comments. Furthermore we are grateful to Marko Robnik for the kind provision of the eigenvalues of the cardioid billiard. We would like to thank the Deutsche Forschungsgemeinschaft for financial support.

# References

[1] G. Casati, B. V. Chirkov and I. Guarneri, Phys. Rev. Lett. **54** (1985) 1350.

[2] M. V. Berry, Proc. R. Soc. London A **400** (1985) 229.

[3] P. M. Bleher and J. L. Lebowitz, J. Stat. Phys. **74** (1994) 167.

[4] S. W. McDonald and A. N. Kaufmann, Phys. Rev. Lett. **42** (1979) 1189.

[5] G. Casati, F. Valz-Gris and I. Guarneri, Lett. Nuovo Cimento **28** (1980) 279.

[6] O. Bohigas, M.-J. Giannoni and C. Schmit, Phys. Rev. Lett. **52** (1984) 1.

[7] M. C. Gutzwiller, *Chaos in classical and quantum mechanics* (Springer, New York, 1990).

[8] R. Aurich and F. Steiner, Physica **D 32**(1988) 451.

[9] E. B. Bogomolny, B. Georgeot, M.-J. Giannoni and C. Schmit, Phys. Rev. Lett. **69** (1992) 1477.

[10] J. Bolte, G. Steil and F. Steiner, Phys. Rev. Lett. **69** (1992) 2188.

[11] R. Aurich, F. Scheffler and F. Steiner, Phys. Rev. E. **51** (1995) 4173.

[12] P. Sarnak, Commun. Math. Phys. **161** (1994) 419.

[13] P. Sarnak, *Arithmetic quantum chaos*, in: *Israel Math. Conference Proceedings* **8**(1995) 183.

[14] F. Steiner, *Quantum chaos,* in *Schlaglichter der Forschung. Zum 75. Jahrestag der Universität Hamburg 1994*, (Ed. R. Ansorge) Festschrift published on the occasion of the 75th anniversary of the University of Hamburg, p.543–564 (Dietrich Reimer Verlag, Berlin und Hamburg, 1994).

[15] R. Aurich, J. Bolte, and F. Steiner, Phys. Rev. Lett. **73** (1994) 1356.

[16] H. P. Baltes and E. R. Hilf, *Spectra of finite systems* (Bibliographisches Institut, Mannheim, Wien, Zürich, 1976).

[17] H.-D. Gräf, H. L. Harney, H. Lengeler, C. H. Lewenkopf, C. Rangacharyulu, A. Richter, P. Schardt and H. A. Weidenmüller, Phys. Rev. Lett. **69** (1992) 1296.

[18] L. A. Bunimovich, Funct. Anal. Appl. **8** (1974) 254.

[19] L. A. Bunimovich, Commun. Math. Phys. **65** (1979) 295.

[20] M. Sieber, U. Smilansky, S. C. Creagh and R. G. Littlejohn, J. Phys. A **16** (1993) 6217.

[21] D. Alonso and P. Gaspard, J. Phys. A: Math. Gen. **27** (1994) 1599.

[22] R. Aurich, T. Hesse and F. Steiner, Phys. Rev. Lett. **79** (1995) 4408.

[23] C. E. Porter (ed.), *Statistical theories of spectra: fluctuations* (Academic Press, New York, 1965).

[24] T. A. Brody, J. Flores, J. B. French, P. A. Mello, A. Pandey and S. S. M. Wong, Rev. Mod. Phys. **53** (1981) 385.

[25] M. L. Mehta, *Random Matrices and the Statistical Theory of Energy Levels* (Academic Press, New York, 1967) and new revised and enlarged edition, 1990.

[26] M. V. Berry and M. Tabor, Proc. R. Soc. Lond. A. **356** (1977) 375.

[27] Z. Cheng and J. L. Lebowitz, Phys. Rev. A **44** (1991) R3399.

[28] A. Bäcker, F. Steiner and P. Stifter, Phys. Rev. E **52**(1995) 2463.

[29] R. Aurich and F. Steiner, Physica **D 82**(1995) 266.

[30] M. V. Berry, Nonlinearity **1** (1988) 399.

[31] R. Aurich and F. Steiner, Physica **D 43**(1990) 155.

[32] M. V. Berry, *Some quantum-to-classical asymptotics*, In *Proceedings of the 1989 Les Houches School on Chaos and Quantum Physics*, 251–303, eds. M.-J. Giannoni, A. Voros, and J. Zinn Justin (Elsevier, Amsterdam, 1991).

[33] M. L. Mehta and J. des Cloizeaux, Ind. J. Pure Appl. Math. **3**(1972) 329.

[34] O. Bohigas and M.-J. Giannoni, in *Mathematical and Computational Methods in Nuclear Physics*, eds. J. S. Dehesa, J. M. G. Gomez and A. Polls, Lecture Notes in Physics Vol. 209, p. 1 (Springer, New York, 1984).

[35] J. B. French, P. A. Mello and A. Pandey, Ann. Phys. (NY) **113**(1978) 277.

[36] A. Pandey, Ann. Phys. (NY) **119**(1979) 170.

[37] M. L. Mehta, Z. Phys. B **86**(1992) 285.

[38] B. Dietz and F. Haake, Z. Phys. B **80**(1990) 153.

[39] E. L. Basor, C. A. Tracy and H. Widom, Phys. Rev. Lett. **69**(1992) 5.

[40] M. M. Fogler and B. I. Shklovski, Phys. Rev. Lett. **74**(1995) 3312.

[41] F. J. Dyson, *The Coulomb fluid and the fifth Painlévé transcendent*, in *Chen Ning Yang*, eds. C. S. Liu and S.-T. Yan (International Press, Boston, 1995).

[42] A. Bäcker, *Spektrale Statistiken des quantisierten Kardioid-Billards*, Diploma Thesis, II. Institut für Theoretische Physik, Universität Hamburg, 1995.

[43] O. Costin and J. L. Lebowitz, Phys. Rev. Lett. **75**(1995) 69.

[44] M. G. Kendall and A. Stuart, *The advanced theory of statistics. Vol. 1: Distribution theory* (Charles Griffin and Company Limited, London, 1963).

[45] R. von Mises, *Mathematical theory of probability and statistics* (Academic Press, New York and London, 1967).

[46] P. M. Bleher, F. J. Dyson and J. L. Lebowitz, Phys. Rev. Lett. **71** (1993) 3047.

[47] P. M. Bleher and J. L. Lebowitz, Ann. Inst. Henri Poincare **31** (1995) 27.

[48] H. Spohn, in *Hydrodynamic behavior and interacting particle systems*, ed. G. Papanicolaou, Lecture Notes from the IMA Vol. 9, p. 151 (Springer, New York, 1987).

[49] P. M. Bleher, Duke Math. J. **67** (1992) 461.

[50] P. M. Bleher, Z. Cheng, F. J. Dyson and J. L. Lebowitz, Commun. Math. Phys. **154** (1993) 433.

[51] J. Bolte, Int. J. Mod. Phys. B **7** (1993) 4451.

[52] A. Selberg, Arch. Math. Naturvid. **48** (1946) 89.

[53] A. Selberg, *Collected Papers Vol. II* (Springer, Heidelberg, 1991), p. 47.

[54] H. L. Montgomery, in: *Number theory, trace formulas and discrete groups*, eds. K. E. Aubert, E. Bombieri and D. Goldfeld (Symposium in Honor of Atle Selberg, Oslo, July 14-21, 1987; Academic Press, 1989).

[55] J. Bolte, talk given at the 59. Physikertagung Berlin 1995.

[56] D. V. Kosygin, A. A. Minasov and Ya. G. Sinai, Russ. Math. Surveys **48**:4 (1993) 1.

[57] P. M. Bleher, *Trace formula for quantum integrable systems, lattice-points problem, and small divisors*, IUPUI Preprint # 96-11, to be published in the Proceedings of the Summer Program on "Emerging Applications of Number Theory", July 15–26 1996.

[58] A. S. Besicovitch, *Almost Periodic Functions* (Dover Pub., Cambridge, 1954)

[59] D. R. Heath-Brown, Acta Arithmetica **60** (1992) 389.

[60] K.-M. Tsang, Proc. London Math. Soc. **65** (1992) 65.

[61] P. M. Bleher and F. J. Dyson, Commun. Math. Phys. **160** (1994) 493.

[62] P. M. Bleher, Duke Math. J. **70** (1993) 655.

[63] P. M. Bleher, Duke Math. J. **74** (1994) 1.

[64] P. M. Bleher, D. V. Kosygin and Ya. G. Sinai, Commun. Math. Phys. **170** (1995) 375.

[65] R. Schubert, *The trace formula and the distribution of eigenvalues of Schrödinger operators on manifolds all of whose geodesics are closed*, DESY report, DESY 95–090, May 1995.

[66] J. Hadamard, J. Math. Pure Appl. **4** (1898) 27.

[67] E. Hopf, Ber. Verh. Sächs. Akad. Wiss. Leipzig **91** (1939) 261.

[68] I. M. Gelfand and S. V. Fomin, Transl. Amer. Math. Soc. **2** No.1 (1955) 49.

[69] Ya. G. Sinai, Sov. Math. Dokl. **2** (1961) 106.

[70] A. Selberg, J. Indian Math. Soc. **20** (1956) 47.

[71] H. Poincaré, Acta Math. **1** (1882) 1.

[72] N. L. Balazs and A. Voros, Phys. Rep. **143** (1986) 109.

[73] R. Aurich and F. Steiner, Physica **D 39** (1989) 169.

[74] O. Bohigas, M. J. Giannoni and C. Schmit, in *Quantum chaos and statistical nuclear physics*, eds. T. H. Seligman and H. Nishioka, Lecture Notes in Physics Vol. 263, p. 18 (Springer, Heidelberg, 1986).

[75] M. Robnik, J. Phys. A: Math. Gen. **16** (1983) 3971.

[76] M. Robnik, J. Phys. A: Math. Gen. **17** (1984) 1049.

[77] M. Wojtkowski, Commun. Math. Phys. **105** (1986) 391.

[78] D. Szász, Commun. Math. Phys. **145** (1992) 595.

[79] R. Markarian, Nonlinearity **6** (1993) 819.

[80] N. I. Chernov and C. Haskell, Ergodic theory and dynamical systems **16** (1996) 19.

[81] H. Bruus and N. D. Whelan, Nonlinearity **9** (1996) 1023.

[82] T. Prosen and M. Robnik, private communication.

[83] T. Prosen and M. Robnik, J. Phys. A: Math. Gen. **26** (1993) 2371.

[84] M. Sieber and F. Steiner, Physica D **44** (1990) 248; Phys. Rev. Lett. **67** (1991) 1941; see also M. Sieber, Ph. D. Thesis, Universität Hamburg, DESY report, DESY 91–030 (1991).

[85] H. Alt, H.-D. Gräf, H. L. Harney, R. Hofferbert, H. Lengeler, C. Rangacharyulu, A. Richter and P. Schardt, Phys. Rev. E **50** (1994) R1.

[86] M. Kac, Am. Math. Monthly **73**(1966) 1.

[87] C. Gordon, D. Webb and S. Wolpert, Bull. Am. Math. Soc. **27**(1992) 134.

[88] S. Sridhar and A. Kudrolli, Phys. Rev. Lett. **72**(1994) 2175.

[89] H. Wu, D. W. L. Sprung and J. Martorell, Phys. Rev. E **51**(1995) 703.

[90] L. Sirko and P. M. Koch, Phys. Rev. E **54**(1996) R21.